\documentclass[journal,twoside,web]{ieeecolor}
\usepackage{generic}
\usepackage{cite}
\usepackage{amsmath,amssymb,amsfonts}
\allowdisplaybreaks
\usepackage{graphicx}
\usepackage{textcomp}
\usepackage{mathtools}
\usepackage{dsfont}
\usepackage{flushend}
\usepackage{enumerate}
\usepackage{dsfont}
\usepackage{pifont}
\usepackage{algpseudocode}
\usepackage{array}
\usepackage{booktabs}
\usepackage{textcomp}
\usepackage{stfloats}
\usepackage{url}
\usepackage{verbatim}
\usepackage{graphicx}
\usepackage[hidelinks]{hyperref}
\usepackage{subcaption}
\usepackage[affil-it]{authblk}
\usepackage{fancyhdr}
\usepackage{bm}
\usepackage[linesnumbered,ruled,vlined]{algorithm2e}
\SetKwInput{KwInput}{Input}
\SetKwInput{KwOutput}{Output}
\SetKw{KwAssert}{Assert}

\usepackage{amsthm}
\newtheorem{lemma}{Lemma}[]
\newtheorem{theorem}{Theorem}[]

\newtheorem{proposition}{Proposition}[]
\newtheorem{corollary}{Corollary}[]
\theoremstyle{definition}
\newtheorem{definition}{Definition}[]
\newtheorem{oldremark}{Remark}[]
\newtheorem{oldexample}{Example}[]

\newtheorem{assumption}{Assumption}[]

\newenvironment{remark}
{\begin{oldremark}\pushQED{\qed}}
	{\popQED\end{oldremark}}
\newenvironment{example}
{\begin{oldexample}\pushQED{\qed}}
	{\popQED\end{oldexample}}

\usepackage{tikz}
\usetikzlibrary{shapes,arrows}
\usetikzlibrary{arrows,calc,positioning}
\tikzset{
	block/.style = {draw, rectangle,
		minimum height=1.2cm,
		minimum width=1.2cm},
	input/.style = {coordinate,node distance=1cm},
	output/.style = {coordinate,node distance=2cm},
	arrow/.style={draw, -latex,node distance=2cm},
	pinstyle/.style = {pin edge={latex-, black,node distance=1cm}},
	sum/.style = {draw, circle, node distance=1cm},
}
\definecolor{backgreen}{HTML}{E9F3DF}
\definecolor{backblue}{HTML}{DAE5EC}
\definecolor{backpurple}{HTML}{E5E0E8}
\definecolor{backorange}{HTML}{F2E9DF}
\definecolor{lightergray}{gray}{0.9}
\definecolor{lightgray}{gray}{0.85}

\usepackage[framemethod=TikZ]{mdframed}
\mdfdefinestyle{callout}{%
	linecolor=black,
	linewidth=1pt,%
	roundcorner=0pt,
	innertopmargin=4pt,
	innerbottommargin=4pt,
	innerrightmargin=5pt,
	innerleftmargin=5pt,
	leftmargin = 0pt,
	rightmargin = 0pt,
	backgroundcolor=lightergray
}

\def\BibTeX{{\rm B\kern-.05em{\sc i\kern-.025em b}\kern-.08em
    T\kern-.1667em\lower.7ex\hbox{E}\kern-.125emX}}
\usepackage{endnotes}

\DeclareMathOperator{\argmin}{\mathrm{argmin}}
\DeclareMathOperator{\argmax}{\mathrm{argmax}}

\DeclareMathOperator{\col}{\mathrm{col}}
\DeclareMathOperator{\diag}{\mathrm{diag}}

\DeclareMathOperator{\sign}{\mathrm{sgn}}
\DeclareMathOperator{\lambdamin}{\lambda_\mathrm{min}}
\DeclareMathOperator{\lambdamax}{\lambda_\mathrm{max}}

\DeclareMathAlphabet{\doublestruck}{U}{BOONDOX-ds}{m}{n}
\newcommand{\ones}[0]{\mathds{1}}
\newcommand{\zeros}[0]{\doublestruck{0}}

\newcommand{\NE}[0]{\mathrm{NE}}

\newcommand{\R}[0]{\mathbb{R}}
\newcommand{\C}[0]{\mathbb{C}}

\newcommand{\Rnn}[0]{\mathbb{R}_{\geq0}}

\newcommand{\Acal}[0]{\mathcal{A}}

\newcommand{\Ccal}[0]{\mathcal{C}}

\newcommand{\Fcal}[0]{\mathcal{F}}

\newcommand{\Kcal}[0]{\mathcal{K}}

\newcommand{\Mcal}[0]{\mathcal{M}}

\newcommand{\Ocal}[0]{\mathcal{O}}
\newcommand{\Pcal}[0]{\mathcal{P}}

\newcommand{\Scal}[0]{\mathcal{S}}

\newcommand{\Ucal}[0]{\mathcal{U}}

\usepackage{xstring}
\newcommand{\weblink}[1]{%
	\StrSubstitute{#1}{https://}{}[\strippedurl]%
	\href{#1}{{\small \color[RGB]{0,0,0} \url{\strippedurl}}}%
}

\newcommand{\BSK}{B_{S\to K}}
\newcommand{\BKS}{B_{K\to S}}
\newcommand{\BSbarK}{B_{\bar{S} \to K}}
\newcommand{\BKSbar}{B_{K\to \bar{S}}}

\newcommand{\Rd}[0]{R_\mathrm{d}}
\newcommand{\Rr}[0]{R_\mathrm{r}}
\newcommand{\Rdbar}[0]{\bar{R}_\mathrm{d}}
\newcommand{\Rrbar}[0]{\bar{R}_\mathrm{r}}
\newcommand{\Rdc}[0]{R_\mathrm{d}^c}
\newcommand{\Rrc}[0]{R_\mathrm{r}^c}

\newcommand{\MSNE}[0]{\mathrm{MSNE}}
\newcommand{\sE}[0]{\mathrm{E}}
\newcommand{\sAE}[0]{\mathrm{AE}}
\newcommand{\sAF}[0]{\mathrm{AF}}
\newcommand{\sF}[0]{\mathrm{F}}

\newcommand{\aN}[0]{0}
\newcommand{\aL}[0]{\mathrm{L}}
\newcommand{\aH}[0]{\mathrm{H}}

\begin{document}
	\title{Evolutionary Dynamics in Continuous-time Finite-state Mean Field Games -- Part~II: Stability}
	\author{Leonardo Pedroso, Andrea Agazzi, W.P.M.H. (Maurice) Heemels, Mauro Salazar
	\thanks{This work was supported in part by the Eindhoven Artificial Intelligence Systems Institute (EAISI).}
	\thanks{L.~Pedroso, W.P.M.H.~Heemels and M.~Salazar are with the Control Systems Technology section, Department of Mechanical Engineering, Eindhoven University of Technology, The Netherlands (e-mail: \{l.pedroso,m.heemels,m.r.u.salazar\}@tue.nl).}
	\thanks{A.~Agazzi is with the Institute of Mathematical Statistics and Actuarial Science, Department of Mathematics and Statistics, University of Bern, Switzerland (e-mail: andrea.agazzi@unibe.ch).}}
	
	\maketitle
	
	\begin{abstract} 
		We study a dynamic game with a large population of players who choose actions from a finite set in continuous time. Each player has a state in a finite state space that evolves stochastically with their actions. A player’s reward depends not only on their own state and action but also on the distribution of states and actions across the population, capturing effects such as congestion in traffic networks. In Part I, we introduced an evolutionary model and a new solution concept -- the mixed stationary Nash Equilibrium (MSNE) -- which coincides with the rest points of the mean field evolutionary model under meaningful families of revision protocols. In this second part, we investigate the evolutionary stability of MSNE. We derive conditions on both the structure of the MSNE and the game’s payoff map that ensure local and global stability under evolutionary dynamics. These results characterize when MSNE can robustly emerge and persist against strategic deviations, thereby providing insight into its long-term viability in large population dynamic games.
	\end{abstract}
	

\section{Introduction}

In Part~I of this work~\cite{PedrosoAgazziEtAl2025MFGAvg}, we show that state-of-the-art solution concepts based on a notion of a behavioral stationary Nash equilibrium do not have an \emph{evolutionary interpretation}. Therein, we define a novel notion of equilibrium for this class of games that does, which we call MSNE. Moreover, for the first time in the literature, we formulate an explicit \emph{mean field evolutionary model of the dynamic game} as an ODE called the master equation
\begin{equation}\label{eq:ODE_mu_ev}
	\dot{\mu}^c[s,u] =f_{s,u}^{c,d}(\mu) + 	f_{s,u}^{c,r}(\mu)  
\end{equation}
$\forall c\in [C]\; \forall s\in \Scal^c \;\forall u\in \Ucal^c_D$, where
\begin{equation*}
	\begin{split}
		f_{s,u}^{c,d}(\mu) =&  \Rdc \sum_{s^\prime \in \Scal^c} \sum_{a^\prime \in \Acal^c(s^\prime)}\!\!\!\phi^c(s|s^\prime,a^\prime)u(a^\prime|s^\prime)\mu^c[s^\prime,u] \\-& \Rdc\mu^c[s,u]
		\end{split}
	\end{equation*}
	\begin{equation}\label{eq:fd_fr}
		\begin{split}
		f_{s,u}^{c,r}(\mu) =& \sum_{u^\prime \in \Ucal^c_D}  \mu^c[s,u^\prime] \rho^c_{u^\prime u}(F^c(\mu),\mu^c[\Scal^c,\cdot]) \\- &\mu^c[s,u] \sum_{u^\prime \in \Ucal_D^c}  \rho^c_{uu^\prime}(F^c(\mu),\mu^c[\Scal^c,\cdot]).
	\end{split}
\end{equation}
The main results of Part~I show, under mild conditions, an equivalence between the MSNE solution concept  and rest points of the evolutionary dynamics \eqref{eq:ODE_mu_ev} for various classes of meaningful revision protocols.

\begin{mdframed}[style=callout]
			Intuitively, if one designs the parameters of a game such that a desired population state $\mu^\star$ is a MSNE, the analysis of Part~I allows to establish mild conditions under which $\mu^\star$ is a rest point of meaningful evolutionary dynamics. However, to guarantee that the population will not diverge away from $\mu^\star$ requires further stability analysis, which we address in this Part~II.
\end{mdframed}

The analysis on the evolutionary stability of MSNE will be divided into three parts, each corresponding to one of the following three sections. First, we establish the instability of non-MSNE rest points of imitative via comparison revision protocols. Second, we establish stability properties when the MSNE is strict, i.e., a MSNE for which a single policy achieves strictly higher payoff than all others for each class. Third, we consider a two time-scale regime whereby the revision dynamics are very slow compared to the state dynamics and discuss dynamic stability of non-strict MSNE.

For the sake of brevity, we reuse all notation from Part~I without stating it here. Similarly, all assumptions introduced in Part~I \cite[Assumptions~1-3]{PedrosoAgazziEtAl2025MFGAvg} are assumed to hold in the analysis of Part~II, regarding differentiability of the single stage reward, uniqueness of recurrent communicating classes of policy Markov chains, and well-posedness of the revision protocols.

\subsection{Preliminaries}	

In this paper, we make use of several notions of stability of an autonomous system of the form
\begin{equation}\label{eq:prelim_autonomous_sys}
	\dot{x} = f(x),
\end{equation}
where $f:D\to \R^n$ is Lipschitz continuous. In what follows, we state the stability definitions that are used throughout the paper. These definitions are stated for a compact set $\Acal \subseteq D$, but one can particularize such definitions for a point $x^\star \in D$ by defining $\Acal = \{x^\star\}$. We say that $\Acal$ is positively invariant w.r.t.\ \eqref{eq:prelim_autonomous_sys} if $x(0)\in \Acal \implies x(t) \in \Acal$ for all $t\geq 0$. If $\Acal = \{x^\star\}$, $\Acal$ being positively invariant is equivalent to $x^\star$ being an equilibrium point of \eqref{eq:prelim_autonomous_sys}. The distance of a point $x\in D$ to $\Acal$ is defined as $d_{\Acal}(x):= \inf_{y\in \Acal} ||x-y||$. We say that $\Acal$ is Lyapunov stable if for all $\epsilon >0$ there is a $\delta>0$ such that $d_\Acal(x(0))< \delta \implies d_\Acal(x(t))<\epsilon\;$ for all $t\geq 0$. We say that $\Acal$ is asymptotically stable if $\Acal$ is Lyapunov stable and there is $\delta >0$ such that $d_\Acal(x(0))< \delta \implies \lim_{t\to \infty} d_\Acal(x(t)) = 0$. We say that $\Acal$ is exponentially stable if there are positive constants $c,k,\lambda >0$ such that $d_\Acal(x(t))\leq kd_\Acal(x(0))e^{-\lambda t}$ for all $d_\Acal(x(0)) < c$. We also say that $\Acal$ is globally asymptotically stable and globally exponentially stable if the respective definitions hold for any $x(0)\in D$. For more details on the stability concepts and stability analysis tools used in this paper see \cite{Isidori1995,Khalil2002}.




\section{Non-MSNE rest points of imitative via comparison revision protocols}\label{sec:unst_nonMSNE_rest_imitative}

Recall that in the equilibrium analysis in Part~I \cite[Theorems~4 and~5]{PedrosoAgazziEtAl2025MFGAvg}, for imitative via comparison revision protocols, a rest point of the evolutionary dynamics is not necessarily a MSNE. The following result shows that non-MSNE rest points are unstable under the evolutionary dynamics \eqref{eq:ODE_mu_ev} even if some (or all) classes use imitative via comparison revision protocols.

\begin{theorem}\label{th:imitative_rest_unstable}
	Consider an imitative via comparison, excess payoff, or pairwise comparison revision protocol $\rho^c$ for each class. Let $\mu^\star$ be a rest point of the evolutionary dynamics \eqref{eq:ODE_mu_ev}. If $\mu^\star$ is not a MSNE, then $\mu^\star$ is not Lyapunov stable under \eqref{eq:ODE_mu_ev} and no solution trajectory of \eqref{eq:ODE_mu_ev} with $\mu(0) \in \mathop{\mathrm{int}}(X)$ converges to $\mu^\star$.
\end{theorem}
\begin{proof}
	See Appendix~\ref{sec:proof_imitative_rest_unstable}.
\end{proof}

The following corollary is the reciprocal of Theorem~\ref{th:imitative_rest_unstable} and allows to conclude that if a trajectory with a non-degenerate initial condition converges to a rest point, the rest point is a MSNE. As a result, under a very weak stability condition, an equivalence can be established between rest points of the evolutionary dynamics and MSNE for imitative via comparison revision protocols.


\begin{corollary}
	Consider an imitative via comparison, excess payoff, or pairwise comparison revision protocol $\rho^c$ for each class $c\in [C]$. If a solution trajectory of \eqref{eq:ODE_mu_ev} with an interior initial condition $\mu(0) \in \mathrm{int}(X)$ converges to $\mu^\star$, then $\mu^\star$ is a MSNE.
\end{corollary}



%
%

\section{Strict MSNE}

In this section, we study the local stability of a MSNE $\mu^\star$ that has the property of having mass on a single policy that achieves a strictly higher payoff than all other policies. Such a MSNE is called a \emph{strict MSNE}, which is formally defined as follows.

\begin{definition}\label{def:strict_MSNE}
	A joint state-policy distribution $\mu \in X$ is said to be a strict MSNE in the average payoff mean field game if $\mu$ is a MSNE and for all $c\in [C]$ and all $u \in \Ucal_D^c$
	\begin{equation*}
		\mu^c[\Scal^c,u] >0 \implies \left( F^c_u(\mu) >  F^c_v(\mu) \quad \forall v \in \Ucal_D^c\setminus \{u\} \right). \tag*{$\triangle$}
	\end{equation*}
\end{definition}

The evolutionary stability analysis for strict MSNE is simpler when compared to a generic MSNE and allows to establish stability results under weaker conditions. Specifically, in the following result, local asymptotic stability of a strict MSNE is established for the three families of deterministic revision protocols.

\begin{theorem}\label{th:stability_strict}
	Consider an imitative, separable excess payoff, or pairwise comparison revision protocol $\rho^c$ for each class $c\in [C]$. Let $\mu^\star$ be a strict MSNE. Then, $\mu^\star$ is locally asymptotically stable under the evolutionary dynamics \eqref{eq:ODE_mu_ev}.
\end{theorem}
\begin{proof}
	See Appendix~\ref{sec:proof_stability_strict}.
\end{proof}

\begin{mdframed}[style=callout]
			Consider that we design the parameters of a game such that a desired population state $\mu^\star$ is a \emph{strict} MSNE. Theorem~\ref{th:stability_strict} guarantees that small perturbations will not make the population state diverge away from $\mu^\star$. Crucially, this is true even if the specific revision model for the players is unknown. We only require that it satisfies the mild meaningful properties of one of the classes of protocols in the conditions of the theorem.
\end{mdframed}

\section{Two Time-scale Stability of non-strict MSNE}\label{sec:tts}

In physically meaningful applications the rate at which the state dynamics evolve is typically significantly faster than the rate of the revision dynamics. One may leverage this observation to carry out a simpler stability analysis of non-strict MSNE in this regime. Specifically, the two time-scale stability analysis of \eqref{eq:ODE_mu_ev} will be performed relying on the single perturbation theory for nonlinear dynamical systems with different time-scales \cite[Chap.~11]{Khalil2002}. The main goal is to find conditions that characterize the stability of $\MSNE(F,\phi)$ under \eqref{eq:ODE_mu_ev} in a regime where  the time scale separation is sufficiently large. Specifically, we are interested in studying a regime where $\Rdbar \gg \Rrbar$ with $\Rdbar = \min_{c\in[C]} \Rdc$ and $\Rrbar= \max_{c\in[C]} \Rrc$. Henceforth, for the sake of simplicity and without any loss of generality, we set $\Rrbar = 1$. First, we provide general results for generic revision protocols. Then, these are used to obtain simple conditions on the payoff structure of the game such that $\mu^\star$ is stable under whole (sub)classes of revision protocols. 



%




Before stating the results, some notation preliminaries are in order. In this section, by abuse of notation, $f^{c,d}(\mu) := \col(f^{c,d}_{s,u}(\mu); s\in \Scal^c, u\in \Ucal^c_D)$, $f^{c,r}(\mu) := \col(f^{c,r}_{s,u}(\mu); s\in \Scal^c, u\in \Ucal_D^c)$ with a concatenation order that is consistent with the concatenation of $\mu^c$, and $f^{r}(\mu) := \col(f^{c,r}(\mu); c\in [C])$ with a concatenation consistent with the concatenation of $\mu$. The concatenation order in these vectors is arbitrary, but it has to be consistent. Henceforth, unless otherwise specified, for each class $c\in [C]$ with $\Scal^c = \{s^c_1, s^c_2, \ldots, s^c_{p^c}\}$ and $\Ucal_D^c = \{u^c_1, u^c_2, \ldots, u^c_{n^c}\}$, we use the ordering convention $(s^c_1, u^c_1), (s^c_2,u^c_1), \ldots, (s^c_{p^c},u^c_{1}), \ldots, (s^c_{p^c},u^c_{n^c})$ for simplicity.

First, notice that $X^c$ is a subset of the vector space $K^c:= \R^{p^cn^c}$, for each class $c\in [C]$, and $X$ is a subset of the vector space $K:= \bigtimes_{c\in [C]}K^c$, where the solution trajectories of \eqref{eq:ODE_mu_ev} are contained. We divide $K$ into two vector subspaces, whose properties allow for an insighful analysis of the dynamics of \eqref{eq:ODE_mu_ev}. For each class $c\in [C]$, define the vector space $S^c\subset K^c$ as the space of state-dynamics-invariant measures, i.e., $S^c:= \{\mu^c\in K^c: Q^c\mu = \zeros\}$, where $Q^c$ is the generator of the Markov chain described by $f^{c,d}$ with a unitary transition rate.
Since $Q^c$ is a linear transformation, $S^c$ is a vector space. By \cite[Assumption~2]{PedrosoAgazziEtAl2025MFGAvg}, the eigenspace of $Q^u$ associated with eigenvalue $\lambda = 0$ has unitary dimension, therefore $\dim(S^c)= n^c$. Let $v^u \in \R^{p^c}$ be the eigenvector of  $Q^{c,u} = (\phi^{c,u}-I)$ associated with eigenvalue $\lambda = 0$ that sums to one, i.e., $\ones^\top v^{c,u} = 1$. Define $\BSK^c \in \R^{p^cn^c \times n^c}$ as
\begin{equation*}
	\BSK^c = \begin{bmatrix}  e^{c,1} \otimes  v^{c,u^c_1} & \cdots & e^{c,n^c} \otimes v^{c,u^c_{n^c}}  \end{bmatrix},
\end{equation*}
whose columns define a basis for $S^c$, where $e^{c,i} \in \R^{n^c}$ has zeros on all entries except for the $i$-th one that is one. We define the complementary subspace of $S^c$, which is denoted as $\bar{S}^c$, as follows. Define $\BSbarK^c \in \R^{p^cn^c \times (p^c-1)n^c}$ as a matrix such that the column space of $[\BSK^c \BSbarK^c]$  is $\R^{p^cn^c}$ and its columns satisfy  $( e^{c,i} \otimes \ones)^\top \BSbarK^c = \zeros^\top$ for all $i\in [n^c]$. Notice that such a matrix $\BSbarK^c$ exists due to the fact that, for each $i\in [n^c]$, $(e_{c,i}  \otimes \ones )$ is orthogonal to all but one column of $\BSK^c$. The vector subspace $\bar{S}^c$ is defined as the column space of $\BSbarK^c$. Notice that $\BSK^c$ and $\BSbarK^c$ are change of basis matrices that map coordinates in a basis of $S^c$ and $\bar{S}^c$, respectively, to Cartesian coordinates in $K^c$. Likewise, we define the linear maps $\BKS^c \in \R^{n^c\times c^cp^c}$ and $\BKSbar^c \in  \R^{(p^c-1)n^c \times p^cn^c}$ as
\begin{equation*}
	\begin{bmatrix}
		\BKS^c \\ \BKSbar^c
	\end{bmatrix} = 
	\begin{bmatrix}
		\BSK^c & \BSbarK^c
	\end{bmatrix}^{-1}.
\end{equation*}
Let $S:= \bigtimes_{c\in [C]} S^c$, $\bar{S}:= \bigtimes_{c\in [C]}  \bar{S}^c$, $\BSK = \diag(\BSK^c, c\in[C])$, $\BSbarK = \diag(\BSbarK^c, c\in[C])$, $\BKS = \diag(\BKS^c, c\in[C])$, and $\BKSbar = \diag(\BKSbar^c, c\in[C])$. Dividing $K$ into $S$ and $\bar{S}$ allows to describe a solution trajectory $\{\mu(t)\}_{t\geq 0}$ of \eqref{eq:ODE_mu_ev} as a trajectory of coordinates in the bases of $S$ and $\bar{S}$. Specifically, one may define, for each class $c\in [C]$,  $x^c(t) = \BKS^c \mu^c(t)$ and $z^c(t) = \BKSbar^c \mu^c(t)$ as coordinates of $\mu^c(t)$ in $S^c$ and $\bar{S}^c$, respectively. Accordingly, one can define $x(t) = \col(x^c(t), c\in[C]) = \BKS \mu(t)$ and $z(t) = \col(z^c(t), c\in[C]) = \BKSbar \mu(t)$ whose time-evolution completely describes the evolution of $\mu(t)$, as established in the following result.

\begin{lemma}\label{lem:ODE_x_z}
	Let $\{x(t),z(t)\}_{t\geq 0}$ be the unique solution with initial conditions $x(0) = \BKS \mu(0)$ and $z(0) = \BKSbar \mu(0)$ to
	\begin{equation}\label{eq:standard_singular_perturbation}
		\begin{split}
			\dot{x}^c_u(t) = &  \sum_{u^\prime \in \Ucal_D^c}  x^c_{u^\prime} \rho^c_{u^\prime u}(F^c(\BSK x + \BSbarK z),x^c) \\
			&\quad - x^c_u \sum_{u^\prime \in \Ucal_D^c} \rho^c_{uu^\prime}(F^c(\BSK x + \BSbarK z),x^c) \\
			\epsilon \dot{z}^c(t) = &  \frac{\Rdc}{\Rdbar}\BKSbar^c Q^c \BSbarK^c z^c\\
			 & \quad + \epsilon\BKSbar^c f^{c,r}(\BSK x + \BSbarK z)
		\end{split}
	\end{equation}
	for all $c\in[C]$ and all $u\in \Ucal_D^c$, where $\epsilon = \Rrbar/\Rdbar$. Then, $\mu(t) = \BSK x(t) + \BSbarK z(t)$ is the unique solution to \eqref{eq:ODE_mu_ev}. Furthermore, $ x^c(t) \in D^c_x := \{x^c\in \Rnn^{n^c}: \ones^\top x^c = m^c\}$ and $z^c(t)\in D^c_z := \{z^c\in \Rnn^{(n^c-1)p^c}: z^c = \BKSbar^c \mu^c \; \text{for some}\; \mu^c \in X^c\}$ for all $t\geq 0$. Accordingly,  $x(t) \in D_x = \bigtimes_{c\in [C]}D_x^c$ and $z(t) \in D_z = \bigtimes_{c\in [C]}D_z^c$ for all $t\geq 0$.
\end{lemma}
\begin{proof}
		See Appendix~\ref{sec:proof_ODE_x_z}.
\end{proof}

\begin{remark}\label{rm:ODE_x_z_interpretation}
	Interestingly, due to the way the change of basis matrix $[\BSK^c \BSbarK^c]$ was defined, \eqref{eq:standard_singular_perturbation} has a very meaningful interpretation: From the proof of Lemma~\ref{lem:ODE_x_z} in Appendix~\ref{sec:proof_ODE_x_z}, specifically Proposition~\ref{prop:prop_S}(vi), we conclude that $\BKS^c = I_{n^c} \otimes  \ones_{p^c}^\top$. It follows immediately that $x^c_u(t) = \mu^c[\Scal^c,u](t)$, i.e., $x^c(t)$ captures the evolution of the mass of class $c$ using each of the policies in $\Ucal_D^c$ (from any state). Conversely, $z(t)$ captures how far away from the stationary distribution the state distribution of the population using each policy is.
\end{remark}

In this section, we analyze (possibly) non-strict MSNE. If a MSNE $\mu^\star$ is strict, then it is isolated, i.e., there is a neighborhood of $\mu^\star$ that does not contain any other MSNE. However, that is not necessarily the case for a non-strict MSNE. For that reason, in this section, we study instead the stability of a closed set of MSNE denoted by $\Mcal \subseteq \MSNE(F,\phi)$. From the equivalence between $\MSNE(F,\phi)$ and the Nash equilibria of the steady-state game, $\NE(\Fcal)$, in \cite[Lemma~3]{PedrosoAgazziEtAl2025MFGAvg}, one can completely characterize $\Mcal$ resorting to an analogous set $\Mcal_{\Fcal} \subseteq \NE(\Fcal)$. Crucially, from the analysis in Remark~\ref{rm:ODE_x_z_interpretation}, $\mu \in \Mcal$ if and only if $x = \BKS \mu \in \Mcal_\Fcal$. 

\begin{mdframed}[style=callout]
	We conclude that the stability of $\Mcal \subseteq \MSNE(F,\phi)$, under the evolutionary dynamics \eqref{eq:ODE_mu_ev}, is completely characterized by the stability of the set $\{(x,z): x\in \Mcal_{\Fcal},\; z = \zeros\}$ under the evolutionary dynamics projected in $S$ and $\bar{S}$ in \eqref{eq:standard_singular_perturbation}.
\end{mdframed}

Furthermore, notice that \eqref{eq:standard_singular_perturbation} is in the standard form of a singular perturbation model \cite[Chap.~11]{Khalil2002}. Indeed, if  $\Rdbar \gg \Rrbar$, then the dynamics of $z(t)$ will be very fast compared with the dynamics of $x(t)$ in \eqref{eq:standard_singular_perturbation}. Intuitively, singular perturbation theory tools allow to conclude on the stability of \eqref{eq:standard_singular_perturbation} by studying the stability of the slow dynamics of $x(t)$ when $z(t)$ is in quasi-steady state, which in this case corresponds to $z(t) \equiv \zeros$ (which follows from solving for $z(t)$ in \eqref{eq:standard_singular_perturbation} after setting $\epsilon = 0$). In this regime, the slow dynamics of $x(t)$ are described by
\begin{equation}\label{eq:tts_red_sys}
	\begin{split}
			\dot{x}^c_u(t)  =&  \sum_{u^\prime \in \Ucal^c_D}  x^c_{u^\prime} \rho^c_{u^\prime u}(\Fcal^c(x),x^c) \\
			 - &  x^c_u \sum_{u^\prime \in \Ucal^c_D} \rho^c_{uu^\prime}(\Fcal^c(x),x^c) \quad  \forall c\in[C] \;\forall u\in \Ucal_D^c,
	\end{split}
\end{equation}
which is called the \emph{reduced model}. Crucially, \eqref{eq:tts_red_sys} defines standard evolutionary dynamics of a static game for the steady-state game.

\begin{mdframed}[style=callout]
	In this section, using singular perturbation theory tools, the goal is to characterize the stability of $\Mcal \subseteq \MSNE(F,\phi)$ under the evolutionary dynamics of the dynamic game $(F,\phi)$ in \eqref{eq:standard_singular_perturbation}, from the stability properties of $\Mcal_{\Fcal} \subseteq \NE(\Fcal)$ under the evolutionary dynamics of the steady-state game $\Fcal$ in  \eqref{eq:tts_red_sys}.
\end{mdframed}

In the following result, in a regime where $\epsilon =\Rrbar/\Rdbar$ is sufficiently close to zero, we establish conditions on the rate of convergence of the reduced model to $\Mcal_{\Fcal}$ to characterize the stability of $\Mcal \subseteq \MSNE(F,\phi)$ under \eqref{eq:ODE_mu_ev}. 

\begin{theorem}\label{th:tts_stability_generic}
	Consider generic revision protocols $\rho^c$ for each class $c\in [C]$ and let $\Mcal_\Fcal \subseteq \NE(\Fcal)$ be closed. If 
	\begin{enumerate}[(i)]
		\item The revision protocols $\rho^c$ satisfy $\mu \in \MSNE(F,\phi) \implies f^{c,r}(\mu) = \zeros\; \forall c\in[C]$;  \label{it:tts_stability_generic_cond_rho}
		\item The evolutionary dynamics of the steady-state game \eqref{eq:tts_red_sys} admit a Lyapunov function $V: \bar{D}_x \to \Rnn$, where $\bar{D}_x \subseteq D_x$ contains a neighborhood of $\Mcal_\Fcal$, that satisfies
		\begin{itemize}
			\item $V$ is continuously differentiable;
			\item $V$ is positive definite, i.e., $V(x) >0\;$ for all $x \in \bar{D}_x\setminus \Mcal_\Fcal$ and $V(x)  = 0\;$ for all $x \in \Mcal_\Fcal$; 
			\item $\dot{V}(x)\leq - \gamma_1 d_{\Mcal_\Fcal}^2(x)$ for all $x \in \bar{D}_x$ for some $\gamma_1>0$;
			\item $||\partial V/\partial x|| \leq \gamma_2 d_{\Mcal_\Fcal}(x)$ for all $x \in \bar{D}_x$ for some $\gamma_2>0$,
		\end{itemize}
		where it is worth recalling that $d_{\Mcal_\Fcal}(x) :=  \inf_{y\in \Mcal_\Fcal} ||x-y||$;
	\end{enumerate}
then there is $\epsilon^\star >0$ such that for all $\epsilon < \epsilon^\star$, $\Mcal \subseteq \MSNE(F,\phi)$ is locally asymptotically stable under the evolutionary dynamics of the dynamic game $(F,\phi)$ in \eqref{eq:ODE_mu_ev}. Furthermore, if $\bar{D}_x = D_x$, then $\Mcal$ is globally asymptotically stable.
\end{theorem}
\begin{proof}
	See Appendix~\ref{sec:proof_tts_stability_generic}.
\end{proof}

Notice that Theorem~\ref{th:tts_stability_generic} establishes asymptotic stability of the dynamic game in a two time-scale regime under two conditions. First, condition~(i) prevents that the flows of the revision dynamics in equilibria move the state distribution away from the equilibrium distribution (which prevents the effect illustrated in \cite[Example~3]{PedrosoAgazziEtAl2025MFGAvg}). By \cite[Lemma~7]{PedrosoAgazziEtAl2025MFGAvg}, all imitative via comparison, excess payoff, and pairwise comparison revision protocols satisfy condition~\eqref{it:tts_stability_generic_cond_rho} of Theorem~\ref{th:tts_stability_generic}.  Second, condition~(ii) characterizes the stability of $\Mcal_\Fcal$ under the evolutionary dynamics of the steady-state game $\Fcal$ in  \eqref{eq:tts_red_sys}. This condition is stronger than local asymptotic stability of $\Mcal_\Fcal$, but weaker than exponential stability of $\Mcal_\Fcal$. As a result, one can state the following corollary of Theorem~\ref{th:tts_stability_generic}.

\begin{corollary}\label{cor:tts_stability_generic_exp}
	Consider imitative via comparison, excess payoff, or pairwise comparison revision protocols $\rho^c$ for each class $c\in[C]$ such that the vector field of the ODE defined by \eqref{eq:tts_red_sys} is continuously differentiable. If  $\Mcal_{\Fcal} \subseteq \NE(\Fcal)$ is locally (globally) exponentially stable under the evolutionary dynamics of the steady-state game $\Fcal$ in \eqref{eq:tts_red_sys}, then there is $\epsilon^\star >0$ such that for all $\epsilon < \epsilon^\star$, $\Mcal \subseteq \MSNE(F,\phi)$ is locally (globally) asymptotically stable under the evolutionary dynamics of the dynamic game $(F,\phi)$ in \eqref{eq:ODE_mu_ev}. 
\end{corollary}
\begin{proof}
	By \cite[Lemma~7]{PedrosoAgazziEtAl2025MFGAvg}, the revision protocols satisfy condition~(i) of Theorem~\ref{th:tts_stability_generic}. Since $\Mcal_{\Fcal} \subseteq \NE(\Fcal)$ is locally (globally) exponentially stable, then there is a local (global) Lyapunov function $V$ that satisfies condition~(ii) of Theorem~\ref{th:tts_stability_generic} by a simple extension of Lyapunov's converse theorem \cite[Theorem~4.14]{Khalil2002} to the stability of a closed set.
\end{proof}

\begin{remark}
	Two technical comments are in order. First, one could state weaker conditions on the Lyapunov function $V$ in Theorem~\ref{th:tts_stability_generic} resorting to bounds of a class $\Kcal$ function of $d_{\Mcal_\Fcal}(x)$ \cite[Definition~4.2]{Khalil2002}. However, those conditions can only be satisfied for \eqref{eq:tts_red_sys} if the class $\Kcal$ function is linear in a neighborhood of zero, which degenerates into the statement of condition~(ii). Second, if one would strengthen condition~(ii) of Theorem~\ref{th:tts_stability_generic} to require exponential stability of $\Mcal_\Fcal$, then one can also establish exponential stability of $\Mcal$ under \eqref{eq:ODE_mu_ev} by making small changes to the proof. As a result, Corollary~\ref{cor:tts_stability_generic_exp} actually establishes exponential stability of $\Mcal$ under \eqref{eq:ODE_mu_ev}.
\end{remark}

Unfortunately, state-of-the-art results on the stability and convergence rate of evolutionary dynamics for the static game \eqref{eq:tts_red_sys} (see \cite{Sandholm2010}) generally do not admit a Lyapunov function that satisfies condition~(ii) of Theorem~\ref{th:tts_stability_generic} for whole (sub)classes of revision protocols. A particular case that allows for that are \emph{imitative via comparison} revision protocols whereby $\Mcal_\Fcal$ is a single point, as established in the following corollary of Theorem~\ref{th:tts_stability_generic}.

\begin{corollary}\label{cor:tts_stability_generic_imm_cmp}
	Let $\Mcal_{\Fcal} = \{x^\star\}$ and define $\Ucal_D^{c\star} = \argmax_{v\in \Ucal_D^c} \Fcal^c_v(x^\star)$ for all $c\in [C]$. Consider any imitative via comparison revision protocols $\rho^c$ for each class $c\in[C]$ such that the vector field of the ODE defined by \eqref{eq:tts_red_sys} is continuously differentiable and its linearization about $x^\star$ has no eigenvalues with zero real parts.  If $u\in\Ucal_D^{c\star} \implies x^{c\star}_u >0$ for all $c\in[C]$ and all $u\in \Ucal_D^c$ and $y^\top D\Fcal(x^\star) y < 0$ for all $y\in \{ w\in TX : u\notin \Ucal_D^{c\star} \implies w^c_u = 0 \;\forall c\in [C] \forall u \in \Ucal_D^c \}$, then there is $\epsilon^\star >0$ such that for all $\epsilon < \epsilon^\star$, the point $\mu^\star  = \BSK x^\star$ is locally asymptotically stable under the evolutionary dynamics of the dynamic game $(F,\phi)$ in \eqref{eq:ODE_mu_ev}. 
\end{corollary}
\begin{proof}
	The corollary is in the conditions of the state-of-the-art result on static games \cite[Theorem~8.5.8]{Sandholm2010}, which establishes exponential stability of $x^\star$ under \eqref{eq:tts_red_sys}. Then, Corollary~\ref{cor:tts_stability_generic_exp} can be used to establish asymptotic stability of $\mu^\star  = \BSK x^\star$ under \eqref{eq:standard_singular_perturbation} in a two time-scale regime. 
\end{proof}

Given that state-of-the-art stability results for the static game generally do not satisfy the conditions of Theorem~\ref{th:tts_stability_generic}, in what follows we relax these conditions.

\begin{theorem}\label{th:tts_ultimate_bd}
	Consider generic revision protocols $\rho^c$ for each class $c\in [C]$ and let $\Mcal_\Fcal \subseteq \NE(\Fcal)$ be closed. Consider that the evolutionary dynamics of the steady-state game \eqref{eq:tts_red_sys} admit a Lyapunov function $V: \bar{D}_x \to \Rnn$, where $\bar{D}_x \subseteq D_x$ contains a neighborhood of $\Mcal_\Fcal$, that satisfies
		\begin{itemize}
			\item $V$ is continuously differentiable and $||\partial V/\partial x||$ is bounded in $\bar{D}_x$;
			\item $V$ is positive definite, i.e., $V(x) >0\;$ for all $x \in \bar{D}_x\setminus \Mcal_\Fcal$ and $V(x)  = 0\;$ for all $x \in \Mcal_\Fcal$; 
			\item  $\dot{V}$ is negative definite, i.e., $\dot{V}(x) < 0\;$ for all $x \in \bar{D}_x\setminus \Mcal_\Fcal$  and  $\dot{V}(x) = 0\;$ for all $x \in \Mcal_\Fcal$.
		\end{itemize}
	Then, for all $B>0$, there exist $\epsilon^\star,B_0,T >0$ such that for all $\epsilon < \epsilon^\star$ trajectories of \eqref{eq:ODE_mu_ev} satisfy
	\begin{equation*}
		d_\Mcal(\mu(0)) \leq B_0 \implies d_\Mcal(\mu(t))\leq B\;\; \forall t \geq T.
	\end{equation*}
	Furthermore, if $\bar{D}_x = D_x$, then for all $B>0$ and all $\mu(0)\in X$ there exist $\epsilon^\star,T >0$ such that for all $\epsilon < \epsilon^\star$ trajectories of \eqref{eq:ODE_mu_ev} satisfy
	\begin{equation*}
		d_\Mcal(\mu(t))\leq B\;\; \forall t \geq T.
	\end{equation*}
\end{theorem}
\begin{proof}
	See Appendix~\ref{sec:proof_tts_ultimate_bd}.
\end{proof}

Theorem~\ref{th:tts_ultimate_bd} establishes that the trajectories of the evolutionary dynamics of the dynamic game $(F,\phi)$ approach an arbitrarily small neighborhood of $\Mcal \subseteq \MSNE(F,\phi)$ for a sufficiently large time-scale separation. This result is weaker than Theorem~\ref{th:tts_stability_generic}, but so are its conditions. Indeed, condition~(i) of Theorem~\ref{th:tts_stability_generic} is no longer required, and condition~(ii) is relaxed to asymptotic stability of $\Mcal_\Fcal$ under the evolutionary dynamics of the static game $\Fcal$ in \eqref{eq:tts_red_sys}.

\begin{remark}
	Two technical remarks are in order. First, Theorem~\ref{th:tts_ultimate_bd} relies on establishing that the norm of the $z$ component of \eqref{eq:standard_singular_perturbation} is ultimately bounded with a bound that is proportional to $\epsilon^\star$. Then, one can write the evolution of the $x$ component of \eqref{eq:standard_singular_perturbation} as the evolutionary dynamics of the steady-state game with a disturbance whose bound is proportional to $||z||$. Finally, resorting to standard input-to-state-stability analysis, one can show that, for sufficiently small $\epsilon$, the trajectories of the dynamic game are close to the trajectories of the static game. Second, even though explicit  regions of attraction are not provided in the statement of the result, these can be easily obtained from the proof.
\end{remark}

\subsection{Global Stability of MSNE}

State-of-the-art results on the stability of the static game do satisfy the conditions of Theorem~\ref{th:tts_ultimate_bd} for whole (sub)classes of revision protocols. In this section, those results are used to characterize the stability of the MSNE of the dynamic game $(F,\phi)$ under the evolutionary dynamics in \eqref{eq:ODE_mu_ev}. First, for a specific class of payoff structures, one can establish global stability results for the whole set of MSNE of the dynamic game as corollaries of Theorem~\ref{th:tts_ultimate_bd} as follows.

\begin{corollary}\label{th:ub_potential_global}
	Consider an excess payoff or pairwise comparison revision protocol $\rho^c$ for each class $c\in [C]$. Consider that the steady-state game $\Fcal$ is a full potential game, i.e., there exists a continuously differentiable function $U:\Rnn^{pn} \to \R$ such that $ \Fcal = \nabla U$, and that $U$ is concave. Then,  $\MSNE(F,\phi)$ is compact and convex and for all $B>0$ and all $\mu(0)\in X$ there exist $\epsilon^\star,T >0$ such that for all $\epsilon < \epsilon^\star$ trajectories of \eqref{eq:ODE_mu_ev} satisfy $d_{\MSNE(F,\phi)}(\mu(t))\leq B\;\;\; \forall t \geq T$. Furthermore, if one considers instead an imitative, excess payoff, or pairwise comparison revision protocol $\rho^c$ for each class $c\in [C]$, the result still holds if $\mu(0) \in X^\star := \{\mu_0\in X: \mu^{c\star}[\Scal^c,u] > 0  \implies \mu^c_0[\Scal,u] > 0, \forall c\in \Ccal_\mathrm{I}\;\forall u\in \Ucal^c_D\; \forall \mu^\star \in \MSNE(F,\phi)\}$, where  $\Ccal_\mathrm{I}$ is the set of indices of the classes that use an imitative revision protocol.
\end{corollary}
\begin{proof}
	See~Appendix~\ref{sec:proof_ub_potential_global}.
\end{proof}

Corollary~\ref{th:ub_potential_global} shows that for the three whole classes of deterministic revision protocols, if the steady-state game admits a concave potential function, then for a sufficiently large time-scale separation, the evolutionary dynamics of the dynamic evolve arbitrarily close to the whole set of MSNE in finite time. By \cite[Lemma~4]{PedrosoAgazziEtAl2025MFGAvg}, games with a nonincreasing rewards congestion game payoff structure are in the conditions of the theorem and so is the motivating application in \cite[Example~1]{PedrosoAgazziEtAl2025MFGAvg}. Notice that Corollary~\ref{th:ub_potential_global} does not hold for imitative dynamics in general when $\mu(0)\in X$. That is due to the fact that if no player initially chooses a policy in the support of $\mu^\star \in \MSNE(F,\phi)$, then that policy will not ever be chosen if revisions rely on the imitation of other players. Nevertheless, inevitable perturbations of the revision protocol in realistic applications make trajectories move away from such degenerate conditions as formally shown in Theorem~\ref{th:imitative_rest_unstable}. 

\begin{corollary}\label{th:ub_stable_global}
	Consider that either all classes use a separable excess payoff revision protocol or all classes use an impartial pairwise comparison revision protocol. Consider that the steady-state game $\Fcal$ is a stable game, i.e., $(y-x)^\top(\Fcal(y)-\Fcal(x))\leq 0$ for all $x,y\in D_x$. Then,  $\MSNE(F,\phi)$ is compact and convex and for all $B>0$ and all $\mu(0)\in D_x$ there exist $\epsilon^\star,T >0$ such that for all $\epsilon < \epsilon^\star$ trajectories of \eqref{eq:ODE_mu_ev} satisfy $d_{\MSNE(F,\phi)}(\mu(t))\leq B\;\;\; \forall t \geq T$.
\end{corollary}
\begin{proof}
	The result follows immediately from noticing the Lyapunov functions defined in \cite[Theorem~7.2.6]{Sandholm2010} and \cite[Theorem~7.2.9]{Sandholm2010} for separable excess payoff and impartial pairwise comparison revision protocols, respectively, are in the conditions of Theorem~\ref{th:tts_ultimate_bd}.
\end{proof}

Notice that the condition of $\Fcal$ being a stable game is weaker that $\Fcal$ admitting a concave potential function, i.e., the latter implies the former. As a result, the conditions on the steady-state game of Corollary~\ref{th:ub_stable_global} are weaker than the ones of Corollary~\ref{th:ub_potential_global}, but Corollary~\ref{th:ub_potential_global} holds for a wider range of revision protocols. 

\subsection{Local Stability of $\Mcal$}

In this section, we weaken the conditions required on the steady-state game $\Fcal$ to characterize the stability of a subset of MSNE of the dynamic game $\Mcal \subseteq \MSNE(F,\phi)$. Specifically, we require only local conditions of  $\Fcal$ to be able to apply the local statement of Theorem~\ref{th:tts_ultimate_bd} to $\Mcal$. 

In this section, we restrict our attention to a particular type of MSNE subset $\Mcal$ that has the particularity of having a \emph{constant payoff vector}. Indeed, notice that the single-stage reward is generally a function of the state-action distribution of the population $\mu_{\Scal \times \Acal}$ or, in particular cases, a function of the action distribution or of the facility usage distribution in case of a congestion game payoff structure. In all these cases, there is a linear map, which we denote by $\Pi$, between a policy distribution $x\in D_x$ and the distribution that shapes the single-stage reward in steady-state.

\begin{example}\label{eg:pi_map}
	In the general case, the map $\Pi_{\Scal \times \Acal}: \R^{n} \to \R^{pq}$ between a policy distribution $x$ and the corresponding steady-state state-action distribution $\mu_{\Scal\times \Acal}$ is described by $\mu^c_{\Scal \times \Acal}[s,a] =  \sum_{u\in \Ucal_D^c}\eta^{c,u}(s)u(a|s)x^c_u \;\;  \forall c\in[C]\; \forall s\in \Scal^c \; \forall a\in \Acal^c.$ Notice that the kernel of $\Pi_{\Scal \times \Acal}$ need not be trivially $\{\zeros\}$.
\end{example}

Even though the map $\Pi_{\Scal \times \Acal}$ described in Example~\ref{eg:pi_map} is suitable for all payoff structures addressed in this paper, the more tailored it is, the weaker are the conditions obtained in what follows. Specifically, the larger $\dim (\ker \Pi)$ the weaker are the stability conditions. Therefore, for example, in applications where the single-stage rewards depend on the action distribution $\mu_\Acal$, the map $\Pi$ should be chosen as the map between a policy distribution $x$ and the corresponding steady-state $\mu_\Acal$. In the following assumption we characterize the sets $\Mcal \subseteq \MSNE(F,\phi)$ under consideration in this section.

\begin{assumption}\label{ass:MF}
  There is $x^\star \in \NE(\Fcal)$ such that $\Mcal_\Fcal$ can be expressed as $\Mcal_\Fcal = \{x\in D_x: x = x^\star + w, \;w\in \ker \Pi \cap \R_{\Ucal_D^\star}\}$, where $\R_{\Ucal_D^\star}:= \{w\in \R^n :  w^c_u = 0 \;\forall c\in[C]\;\forall u\notin \Ucal_D^{c\star}\}$ and $\Ucal_D^{c\star} = \argmax_{v\in \Ucal^c_D} \Fcal^c_v(x^\star)$ for all $c\in [C]$.
  
\end{assumption}

Under Assumption~\ref{ass:MF}, we consider sets that, given any $x^\star \in \NE(\Fcal)$, extend in directions that only place mass on payoff maximizing policies, i.e., along $\R_{\Ucal_D^\star}$, and that keep the steady-state payoff constant, i.e., along $\ker \Pi$. Define $n^{c\star} := |\Ucal_D^{c\star}|$ and $n^\star := \sum_{c\in[C]}n^{c\star}$. The following lemma establishes properties of $\Mcal_\Fcal$ under Assumption~\ref{ass:MF}.

\begin{lemma}\label{lem:MF}
	Let $\Mcal_\Fcal$ satisfy Assumption~\ref{ass:MF}. Then: (i)~$\Mcal_\Fcal$ is compact and convex; (ii)~$\Fcal(x) = \Fcal(y)$ for all $x,y \in \Mcal_\Fcal$; (iii)~$\Mcal_\Fcal \subseteq \NE(\Fcal)$; and (iv)~$D\Fcal(x) = D\Fcal(y)$ for all $x,y \in \Mcal_\Fcal$.
\end{lemma}
\begin{proof}
	See Appendix~\ref{sec:proof_lem_MF}.
\end{proof}

In what follows, we introduce the conditions on $\Fcal$ about $\Mcal_\Fcal$ that will be required to establish stability results. If $\Mcal_\Fcal$ satisfies those conditions it is said to be a \emph{regular evolutionarily stable set} (ESS), which is defined as follows.

\begin{definition}\label{def:RESS}
	A set $\Mcal_\Fcal$ that satisfies Assumption~\ref{ass:MF} is said to be a regular ESS if there is $x\in \Mcal_\Fcal$ such that $x^c_u>0$ for all $c\in [C]$ and all $u\in \Ucal_D^{c\star}$; and
	\begin{equation}\label{eq:ESS_diff_cond}
		w^\top D\Fcal(x^\star) w < 0, \quad \forall w \in  TX \cap  \R_{\Ucal_D^\star}\setminus  \ker \Pi.
	\end{equation}
	In that case, the corresponding set $\Mcal \subseteq D_x$ is also said to be a regular ESS. \hfill $\triangle$
\end{definition}

The definition of a regular ESS has two conditions. The first states that unused policies in $\Mcal_\Fcal$ have a strictly lower payoff than policies used in $\Mcal_\Fcal$. The second condition, enforces local stability of $\Fcal$ against mutations in payoff maximizing policies towards the outside of $\Mcal_\Fcal$. Given a know MSNE $\mu^\star$ (that, for example, one prescribed to be an equilibrium) an important question is how to assess whether $\Mcal$ is a regular ESS. First, one needs to check whether payoff maximizing policies at $x^\star = \BKS\mu^\star$ are used either in $x^\star$ or along directions of $\R_{\Ucal_D^\star} \cap  \ker \Pi$. Second, to check whether \eqref{eq:ESS_diff_cond} holds, one can use the following proposition to easily conclude on that by computing the eigenvalues and eigenvectors of  a low dimension symmetric matrix.

\begin{proposition}\label{prop:check_ESS}
	Consider any $x\in \Mcal_\Fcal$, define $\Phi$ as a matrix whose columns form an orthonormal basis for $TX \cap  \R_{\Ucal_D^\star}$, and define $\Phi^\perp$ as a matrix such that the columns of $[\Phi \; \Phi^\perp]$ form an orthonormal basis for $\R^{n}$. Condition \eqref{eq:ESS_diff_cond} holds if and only if all eigenvalues of $G +G^\top$, with $G = \Phi^\top D_y \Fcal(\Phi y +\Phi^\perp(\Phi^\perp)^\top x^\star)\vert_{y = \Phi^\top x^\star}$, are nonpositive and every real eigenvector $v \in \R^{n^\star-C}$ in the eigenspace associated with the null eigenvalue satisfies $\Pi \Phi v = \zeros$, i.e., $\Phi v \in \ker \Pi$.
\end{proposition}
\begin{proof}
	See Appendix~\ref{sec:proof_prop_check_ESS}.
\end{proof}

\begin{remark}
	The fact that the revision dynamics are defined over policies may allow for directions of perturbations to a NE policy distribution that place mass in payoff maximizing policies and preserve the payoff vector. This was also observed in the framework proposed in \cite{AltmanHayel2010}, which required introducing a notion of equivalent policies therein. In classical static games, such directions generally do not exist and local stability is studied w.r.t.\ to a single point that is an isolated NE \cite[Chap.~8]{Sandholm2010}. In this paper, an extension of these results to a set $\Mcal_\Fcal$ rather than a single point $\{x^\star\}$ has to be applied to use Theorem~\ref{th:tts_ultimate_bd} under meaningful conditions. Interestingly, notice that particularizing Definition~\ref{def:RESS} to the case where $\Mcal_\Fcal = \{x^\star\}$ degenerates into the state-of-the-art definition of a regular evolutionarily stable state $x^\star$ \cite[Chap.~8.3]{Sandholm2010}.
\end{remark}

\begin{lemma}\label{lem:ESS}
	Let $\Mcal_\Fcal$ be a regular ESS. Then, $\Mcal_\Fcal$ is an isolated set of NE, i.e., there is neighborhood $\Ocal$ of $\Mcal_\Fcal$ where $x\in \Ocal \cap \NE(\Fcal)$ if and only if $x\in \Mcal_\Fcal$. Furthermore, there is a neighborhood of $\Ocal$ of $\Mcal_\Fcal$ where $w^\top D\Fcal(x) w^\top \leq 0$ for all $x\in \Ocal$ and for all $w\in  TX \cap  \R_{\Ucal_D^\star}$ with equality if and only if $w \in \ker \Pi$.
\end{lemma}
\begin{proof}
	See Appendix~\ref{sec:proof_lem_ESS}.
\end{proof}

The properties of Lemma~\ref{lem:ESS} allow to apply state-of-the-art results to show that a regular ESS $\Mcal_\Fcal$ is in the conditions of Theorem~\ref{th:tts_ultimate_bd}.

\begin{corollary}\label{cor:tts_ub_local_ESS}
	Consider that either all classes use a separable excess payoff revision protocol or all classes use an impartial pairwise comparison revision protocol. If $\Mcal_\Fcal$ is a regular ESS, then for all $B>0$ there exist $\epsilon^\star,B_0,T >0$ such that for all $\epsilon < \epsilon^\star$ trajectories of \eqref{eq:ODE_mu_ev} satisfy $d_\Mcal(\mu(0)) \leq B_0 \implies d_\Mcal(\mu(t))\leq B\;\; \forall t \geq T$.
\end{corollary}
\begin{proof}
	Given the properties of Lemma~\ref{lem:ESS} for a regular ESS $\Mcal_\Fcal$, the Lyapunov functions defined in \cite[Theorem~8.4.7]{Sandholm2010} satisfy the conditions of the local version of Theorem~\ref{th:tts_ultimate_bd}.
\end{proof}

Corollary~\ref{cor:tts_ub_local_ESS} is quite general. Under a local condition on the payoff of the steady-state game $\Fcal$ (that is easy to check by Proposition~\ref{prop:check_ESS}) and for a sufficiently large time-scale separation, trajectories of the evolutionary dynamics of the dynamic game approach the set $\Mcal \subseteq \MSNE(F,\phi)$ if initialized sufficiently close. 

\begin{oldexample}
In this example, we illustrate and compare how trajectories of the mean field evolutionary dynamics behave about a ESS and a non-ESS MSNE set $\Fcal$. For that purpose, we generate two single-class random game models that satisfy the modeling assumptions of this paper. Both dynamic games admit MSNE sets $\Mcal$ of dimension one, both admit four payoff maximizing policies, but one MSNE set is a regular ESS and the other is not. On the one hand, in Fig.~\ref{fig:eg_MF_ESS} we depict the regular ESS MSNE set $\Mcal_\Fcal$ along two components of $x$ that correspond to payoff maximizing policies. We also depict several trajectories of the evolutionary dynamics \eqref{eq:ODE_mu_ev}, each with a different random initial condition. Even though the regular ESS condition only provides local stability guarantees, one concludes from Fig.~\ref{fig:eg_MF_ESS} that all trajectories eventually approach $\Mcal$. On the other hand, in Fig.~\ref{fig:eg_MF_nonESS} we depict the non-ESS MSNE set $\Mcal_\Fcal$ along two components of $x$ that also correspond to payoff maximizing policies. In this case, no matter how close to $\Mcal$ the trajectories are initialized and no matter how large $\Rd/\Rr$ is, the trajectories diverge away from $\Mcal$ to other MSNE. All the code used to generate this example is available in an open-access repository at \weblink{https://github.com/fish-tue/evolutionary-mfg-avg}. Therein, more plots of all components of the trajectories shown in Figs.~\ref{fig:eg_MF_ESS} and~\ref{fig:eg_MF_nonESS} can be seen and more systems can be seamlessly analyzed. \hfill $\triangle$

\begin{figure}[t]
	\centering
	\includegraphics[width=0.9\linewidth]{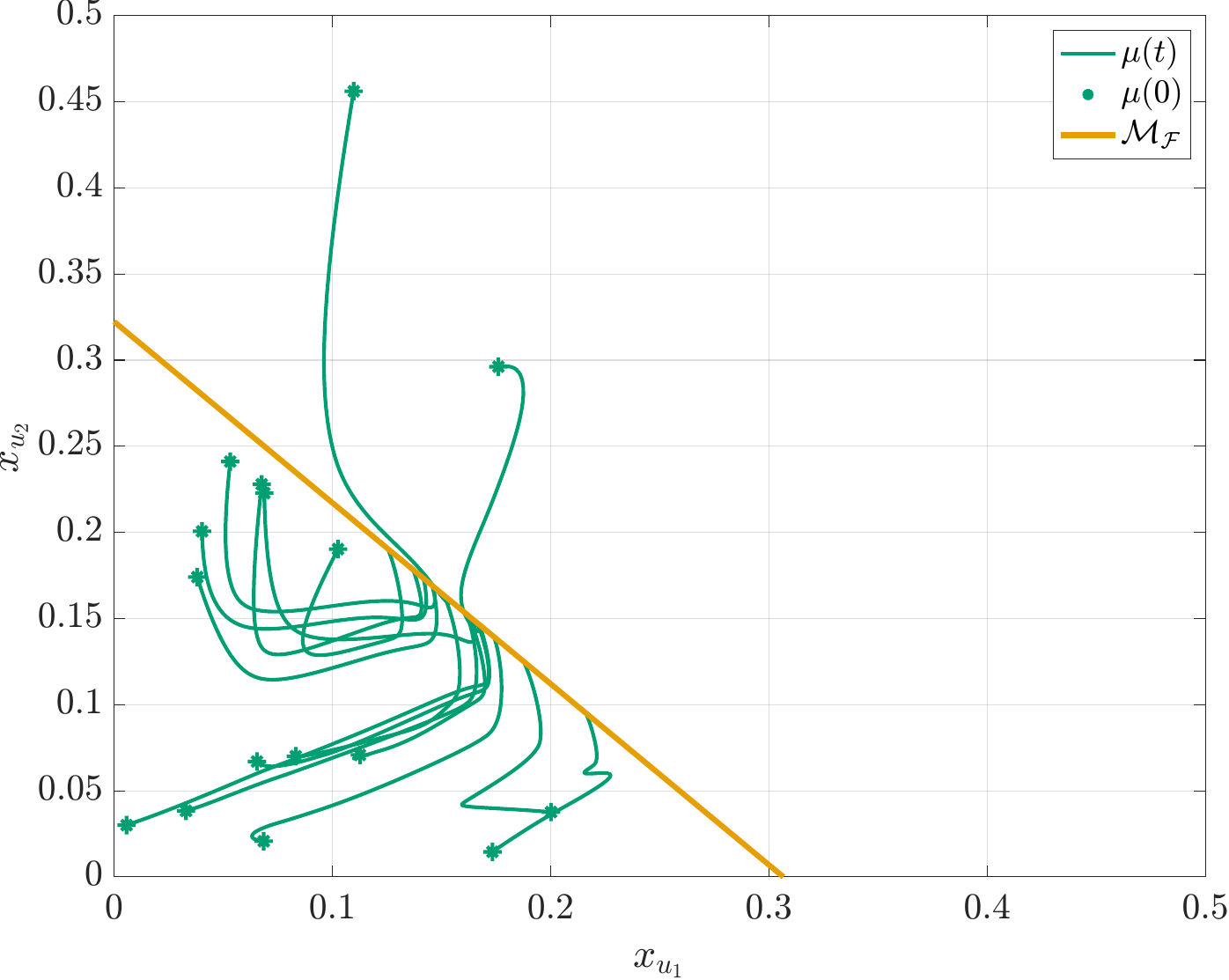}
	\caption{Trajectories of mass on two policies about a regular ESS $\Mcal_\Fcal$.}
	\label{fig:eg_MF_ESS}		
\end{figure}

\begin{figure}
	\centering
	\includegraphics[width=0.9\linewidth]{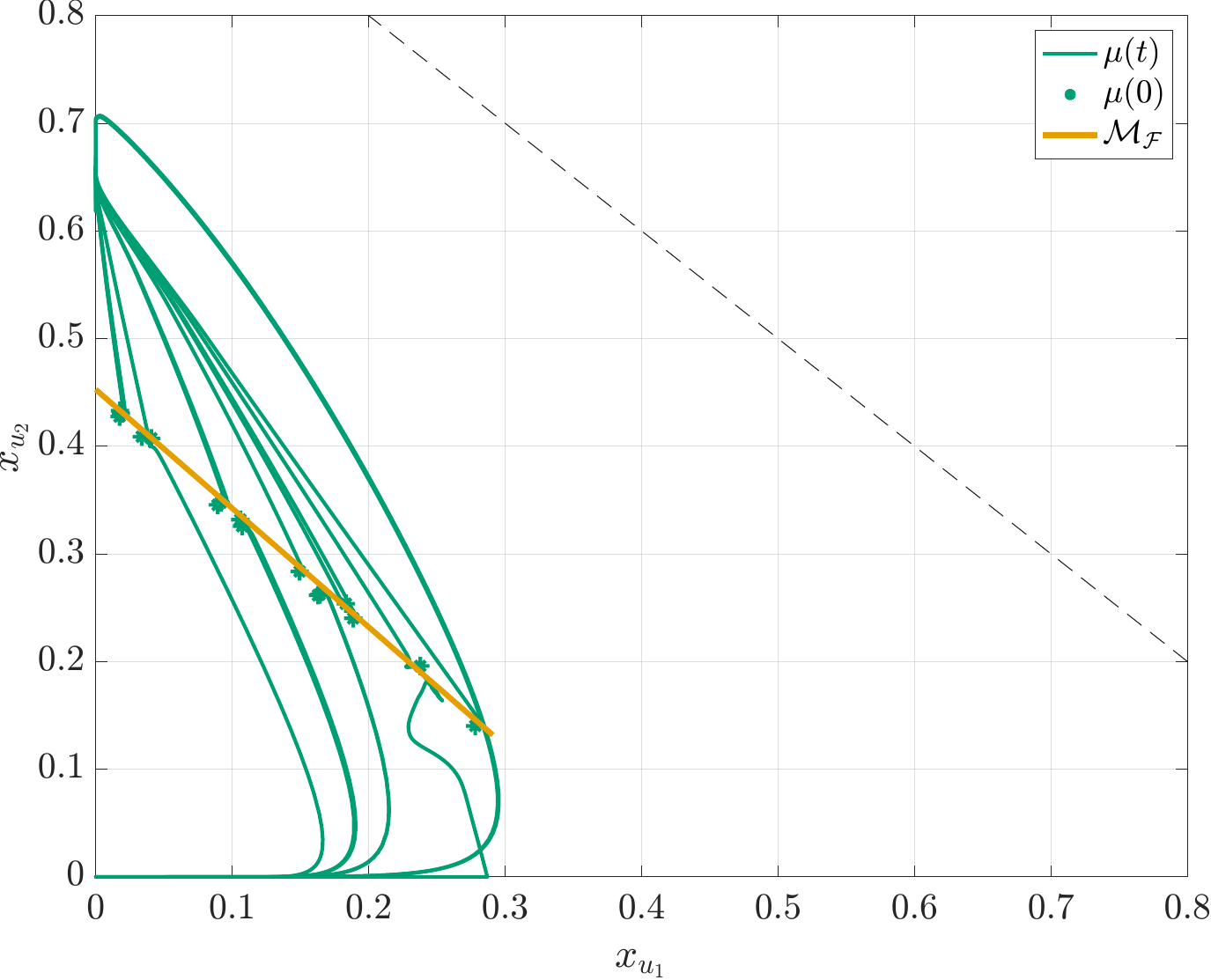}
	\caption{Trajectories of mass on two policies about a non-ESS $\Mcal_\Fcal$.}
	\label{fig:eg_MF_nonESS}
\end{figure}

\end{oldexample}

\section{Medium Access Game: Stability}\label{sec:MAC}

In this section, we illustrate the notions of stability of MSNE resorting to a simple real-life application of a medium access game (MAC) between mobile terminals competing for a common wireless channel similarly to Part~I \cite[Section~VI]{PedrosoAgazziEtAl2025MFGAvg}.

\subsection{Model}

To illustrate more complex behavior we add one more battery state w.r.t.\ the model presented in Part~I. For the sake of completeness, the model is formally characterized by: 
	\begin{itemize}
		\item \emph{Time}: Each player makes a decision each time a Poisson clock with rate $\Rd$ rings. 
		\item \emph{States}: There are four states $\Scal=\{\sE, \sAE,\sAF,\sF\}$ corresponding to four battery levels.
		\item\emph{Actions}: There are three actions $\Acal = \{\aN,\aL,\aH\}$ corresponding to not transmitting, transmitting at low power, and transmitting at high power. We choose $\Acal(\sE) = \{0\}$, $\Acal(\sAE) = \{\aL\}$, $\Acal(\sAF) = \{\aL,\aH\}$, and $\Acal(\sF) = \{\aL,\aH\}$. The transmission powers of actions $\aN$, $\aL$, and $\aH$ are denoted respectively by $P_\aN = 0$, $P_\aL$ and $P_\aH$, which satisfy $0<P_\aL<P_\aH$.
		\item\emph{State transitions}: When a player takes action $\aN$ in state $\sE$ the battery level will be recharged and transition to state $\sF$ with probability $p_\sF$ and will remain at $\sE$ with probability $1-p_\sF$. When a player plays $a \in \{\aL,\aH\}$, the probability of transitioning to the next lower battery state is $\alpha P_a +\gamma$ and of staying in the same energy level is $1-\alpha P_a -\gamma$. Here, $\alpha>0$ and $\gamma>0$ are constants that model the energy consumption due to the transmission of the message and due to other activities, respectively. These constants must satisfy $\alpha P_\aH + \gamma \leq 1$.
		\item \emph{Single-stage reward}: The single-stage reward of a player in state $s$ playing action $a$ when the state-action distribution of the population is $\mu_{\Scal\times \Acal}\in X_{\Scal\times \Acal}$ is the expected signal to interference and noise ratio given by
		\begin{equation*}
			r(s,a,\mu_{\Scal \times \Acal}) \!=\! \frac{P_a}{\sigma^2 + \Rd TC\!\!\!\!\!\sum\limits_{a^\prime \in \{\aL,\aH\}}\!\!\!\!\! P_{a^\prime}\mu_{\Scal \times \Acal}[\Scal,a^\prime]} -\beta P_a,
		\end{equation*}
		where $\sigma, C$, and $\beta$ are constants whose physical interpretation is described in \cite{WiecekAltmanEtAl2011}, and $T$ is the duration of the transmission of a message.
\end{itemize}

There are four deterministic policies, which we denote by $\Ucal_D = \{u_1,u_2,u_2,u_4\}$. These policies are characterized by 
\begin{equation}\label{eq:mac_policies}
	\small\begin{split}
		u_1(\sE) &= u_2(\sE) = u_3(\sE) = u_4(\sE) = \delta_{\aN}(a)\\
		u_1(\sAE) &= u_2(\sAE) = u_3(\sAE) = u_4(\sAE) = \delta_{\aL}(a)\\
		u_1(\sAF) &= u_2(\sAF) = \delta_{\aL}(a), \quad  u_3(\sAF) = u_4(\sAF) = \delta_{\aH}(a)\\
		u_1(\sF) &= u_3(\sF) = \delta_{\aL}(a), \quad  u_2(\sF) = u_4(\sF) = \delta_{\aH}(a).
	\end{split}	
\end{equation}
We denote the unique steady-state distribution of the policies $u_1,u_2,u_3,$ and $u_4$ by $\eta^{u_1},\eta^{u_2},\eta^{u_3},$ and $\eta^{u_4}$, respectively.

\subsection{Stability of MSNE}

To compute a MSNE we solve $\dot{x}(t) = 0$ and $\dot{z}(t) = 0$ in \eqref{eq:standard_singular_perturbation} for a pairwise comparison revision protocol. The solution $x^\star$ and $z^\star = \zeros$ can be used to compute $\mu^\star = \BSK x^\star$. A MSNE for this game exists since it satisfies the conditions of \cite[Theorem~1]{PedrosoAgazziEtAl2025MFGAvg}, so by \cite[Theorem~4]{PedrosoAgazziEtAl2025MFGAvg} a rest point of the evolutionary dynamics under a pairwise comparison protocol exists. Furthermore, by \cite[Theorem~5]{PedrosoAgazziEtAl2025MFGAvg} a rest point is a MSNE under a pairwise comparison revision protocol. Therefore, one concludes that a solution $x^\star$ and $z^\star = \zeros$ exists and that $\mu^\star = \BSK x^\star$ is a MSNE. To define $\Mcal$, one first needs to define the map $\Pi$ between a policy distribution $x$ and the corresponding steady-state action distribution $\mu_\Acal$. The matrix representation of the map $\Pi$ follows from \eqref{eq:mac_policies} and is represented in matrix form in \eqref{eq:PI_too_long} below.
\begin{figure*}[!t]
\begin{equation}\label{eq:PI_too_long}
	\Pi = \begin{bmatrix}
		\eta^{u_1}(\sE) & \eta^{u_2}(\sE) & \eta^{u_3}(\sE) & \eta^{u_4}(\sE)\\
		\eta^{u_1}(\sAE) + \eta^{u_1}(\sAF) + \eta^{u_1}(\sF)& \eta^{u_2}(\sAE)+\eta^{u_2}(\sAF) & \eta^{u_3}(\sAE) + \eta^{u_3}(\sF) & \eta^{u_4}(\sAE)\\
		0 &  \eta^{u_2}(\sF) &  \eta^{u_3}(\sAF) &  \eta^{u_4}(\sAF) +  \eta^{u_4}(\sF)
	\end{bmatrix}
\end{equation}
\end{figure*}

For randomly generated parameters we obtain, rounded to two decimal places,
\begin{equation*}
	\BSK = \diag \left( \begin{bmatrix} 0.07 \\  0.31 \\ 0.31 \\ 0.31  \end{bmatrix}, \begin{bmatrix} 0.09 \\    0.42 \\   0.42 \\   0.06   \end{bmatrix}, \begin{bmatrix} 0.09 \\    0.42 \\    0.06 \\    0.42  \end{bmatrix}, \begin{bmatrix} 0.15 \\    0.67 \\    0.09 \\    0.09 \end{bmatrix}\right) 
\end{equation*}
and  $x^\star = [0.87 \;0.03 \; 0.03 \;0.08]^\top.$
We conclude that the four policies are payoff maximizing. Following the notation of Assumption~\ref{ass:MF}, $\Mcal_\Fcal$ is characterized by $x^\star$ and the kernel of $\Pi$ as $\Mcal_\Fcal = \{x\in D_x :x = x^\star + \theta_1 [ {-0.37} \;   {-0.32} \;    0.86 \;   {-0.17} ]^\top  + \theta_2 [0.55 \;   {-0.80} \;  {-0.01}  \; 0.26 ]^\top, \theta_1,\theta_2\in \R \}.$
To assess whether $\Mcal_\Fcal$ (and, as a result, $\Mcal$) is a regular ESS according to Definition~\ref{def:RESS}, one condition is that unused policies in $\Mcal_\Fcal$ have strictly lower payoff than policies used in $\Mcal_\Fcal$, which is the case in this example. To check the other condition we use Proposition~\ref{prop:check_ESS}. We obtain
\begin{equation*}
	G +G^\top = \underbrace{\begin{bmatrix}
		0.06 &   -1.00 &  -0.07 \\
		0.06 &     0.07 &  -1.00 \\
		1.00 &    0.06 &   0.07
	\end{bmatrix}}_{V:=}\diag(-0.23,0,0)V^\top.
\end{equation*}
Since $ \Phi [-1.00 \;\; 0.07\;\; 0.06]^\top$ and $\Phi [-0.07 \;\; -\!1.00 \;\; 0.07]^\top$ are in the kernel of $\Pi$, it follows from  Proposition~\ref{prop:check_ESS} that $\Mcal_\Fcal$ and $\Mcal$ are regular ESS. Notice that, even though $\mu$ has $16$ components, the stability condition relies on an eigendecomposition of a $3\times 3$ matrix.

\begin{figure*}[t]
	\centering
	\begin{subfigure}{0.47\textwidth}
		\centering
		\includegraphics[width=0.9\textwidth]{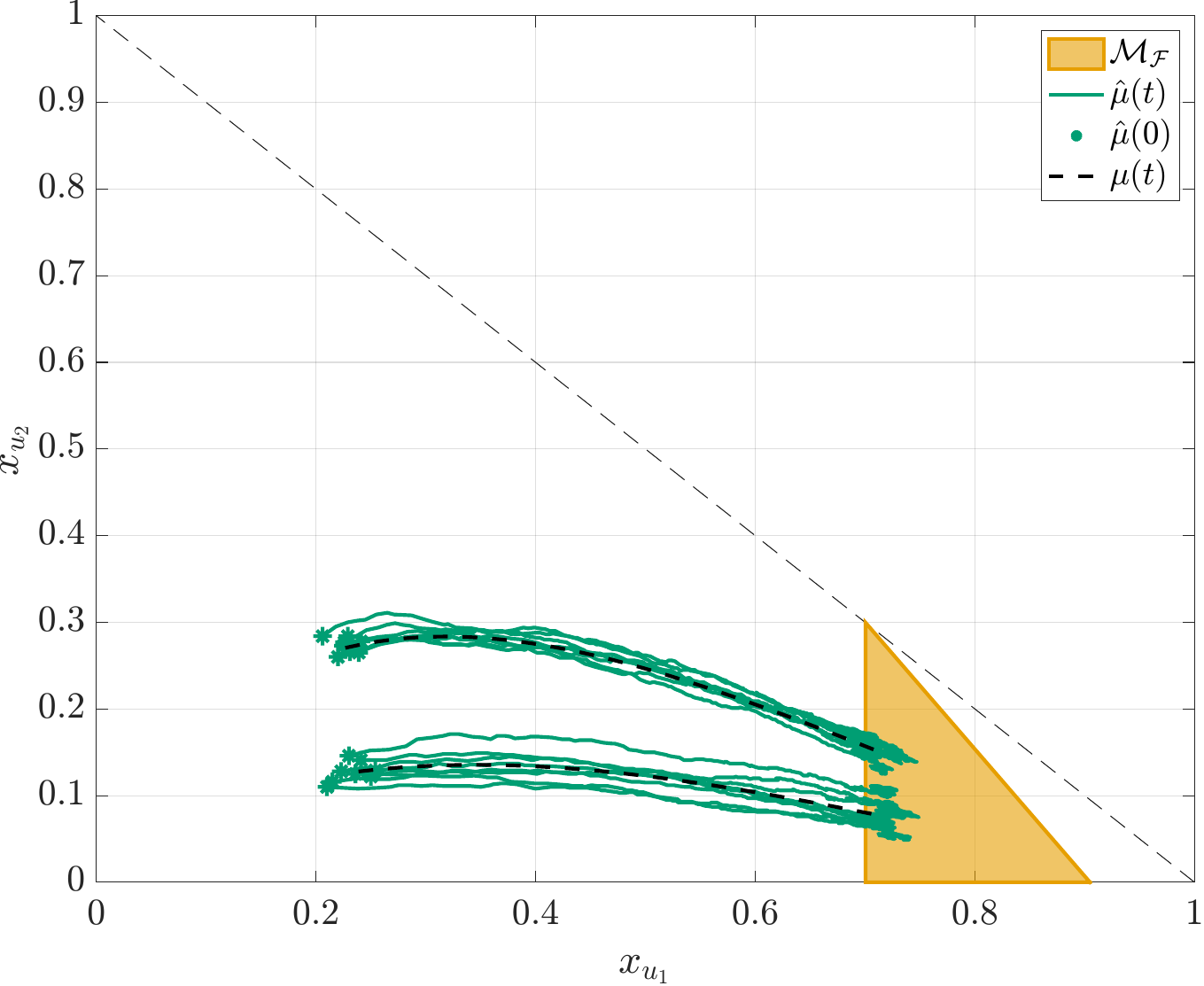}
		\caption{Trajectories of $x_{u_1}$ and $x_{u_2}$ according to \eqref{eq:standard_singular_perturbation}.}
		\label{fig:eg_mac_ESS12}
	\end{subfigure}%
	\hfill%
	\begin{subfigure}{0.47\textwidth}
		\centering
		\centering
		\includegraphics[width=0.9\textwidth]{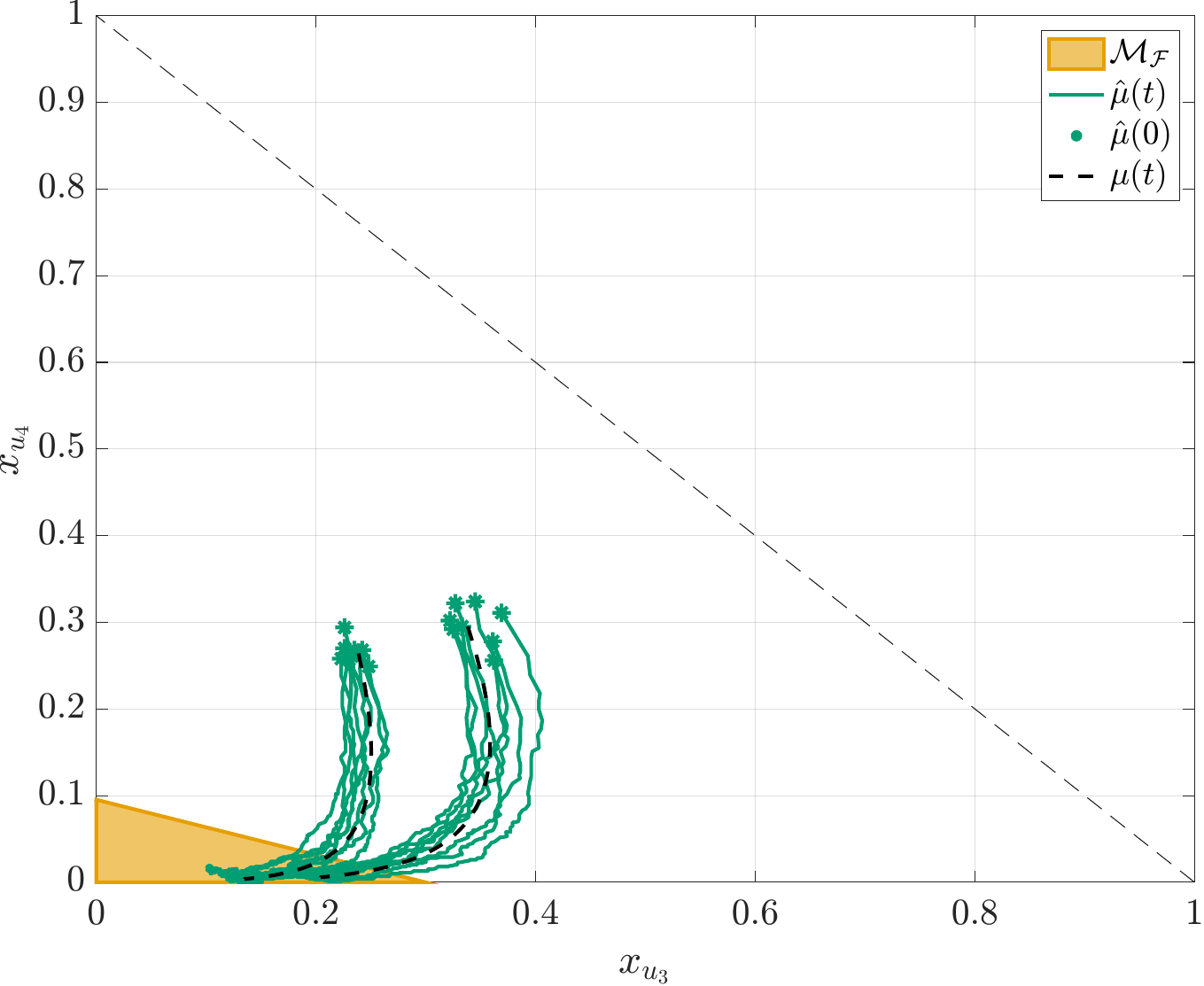}
		\caption{Trajectories of $x_{u_3}$ and $x_{u_4}$ according to \eqref{eq:standard_singular_perturbation}.}
		\label{fig:eg_mac_ESS34}
	\end{subfigure}\\ \vspace{0.2cm}
	\begin{subfigure}{0.47\textwidth}
		\centering
		\centering
		\includegraphics[width=0.95\textwidth]{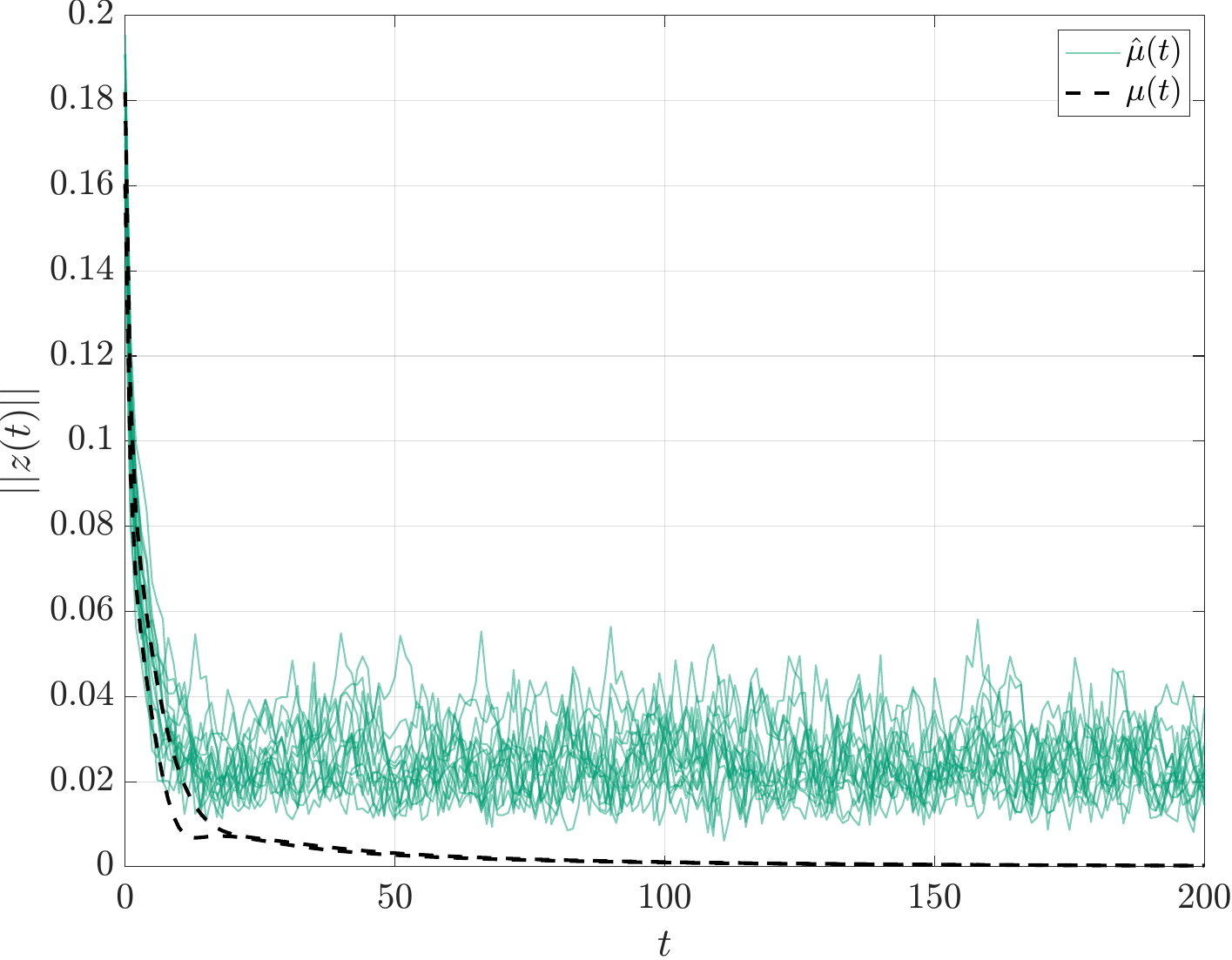}
		\caption{Evolution of the norm of $z$ according to \eqref{eq:standard_singular_perturbation}.}
		\label{fig:eg_mac_ESSz}
	\end{subfigure}%
	\hfill%
	\begin{subfigure}{0.47\textwidth}
		\centering
		\centering
		\includegraphics[width=0.95\textwidth]{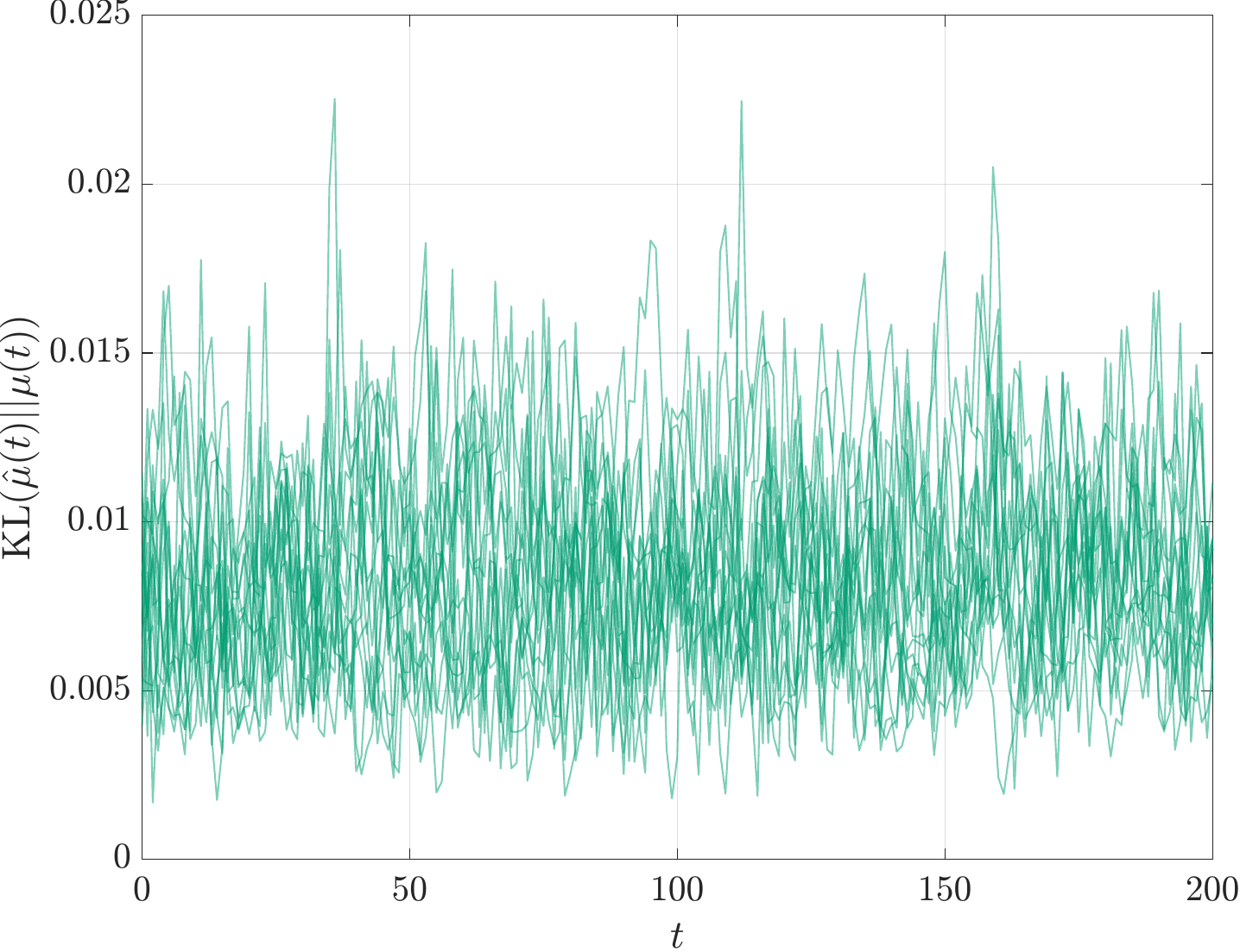}
		\caption{KL divergence between finite and mean field trajectories.}
		\label{fig:eg_mac_ESS_kl}
	\end{subfigure}
	\caption{Trajectories of a finite-population in the illustrative MAC game.}
	\label{fig:eg_mac_ESS}
\end{figure*}

In Fig.~\ref{fig:eg_mac_ESS} we simulate two mean field trajectories with different initial conditions. Moreover, we also simulate several finite-population trajectories for each mean field trajectory, where the player's initial states and policies are drawn randomly from the initial condition of the corresponding mean field trajectory. The finite-population trajectories are simulated with $10^3$ players. Specifically, Figs.~\ref{fig:eg_mac_ESS12} and~\ref{fig:eg_mac_ESS34} depict the set $\Mcal_\Fcal$ and trajectories along the marginal policy distribution components. Fig.~\ref{fig:eg_mac_ESSz} depicts the evolution of the norm of the $z$ component of the trajectories according to \eqref{eq:standard_singular_perturbation}. Fig.~\ref{fig:eg_mac_ESS_kl} plots the difference between the mean field and the finite-population trajectories resorting to the Kullback–Leibler divergence metric. First, we notice from Figs.~\ref{fig:eg_mac_ESS12},~\ref{fig:eg_mac_ESS34}, and~\ref{fig:eg_mac_ESSz} that all trajectories approach $\Mcal$. Second, notice that although the mean field model does converge to $\Mcal$, the finite-population trajectories approach the boundary of $\Mcal$ and the fluctuate about it. Third, we can also conclude from Fig.~\ref{fig:eg_mac_ESS_kl} that the mean field model is a good approximation for the finite-population model as expected from \cite[Theorem~3]{PedrosoAgazziEtAl2025MFGAvg}.

All the code used to generate this example is available in an open-access repository at \weblink{https://github.com/fish-tue/evolutionary-mfg-avg}. Therein, the MAC game is also numerically analyzed for a generic number of battery and transmission power levels.


\section{Conclusion}\label{sec:conclusion}

In Part~II of this work, we study the evolutionary stability of mixed stationary Nash equilibria (MSNE). First, we strengthen the results of Part~I by showing that non-MSNE rest points are unstable under evolutionary dynamics even if some (or all classes) use imitative via comparison revision protocols. Second, we show that if a MSNE is strict, i.e., each class places mass on a single policy with strictly higher payoff than all others, then the MSNE is locally asymptotically stable under meaningful revision protocols. Third, in a two time-scale regime where state dynamics  are significantly faster than revision dynamics, we conclude that there are two classes of games (potential and stable games) for which all trajectories approach the whole set of MSNE. Fourth, in the two time-scale regime, we establish that a simple condition on the payoff structure of the game guarantees that all trajectories initialized sufficiently close to a subset of MSNE remain close to it. 

All in all, a solution concept grounded in an evolutionary model offers a more compelling notion of a game's outcome than one that is not, such as the NE, because it offers insight into how the outcome emerges under very limited assumptions about the players' knowledge. In this work, we abandon state-of-the-art solution concepts for the class of continuous-time finite-state stochastic dynamic games of many players, which do not have an evolutionary interpretation. Instead, we propose a novel solution concept that does, which we call MSNE. This is significant from a \emph{design perspective}. Indeed, one can design a dynamic game such that a desired population state is a MSNE. The tools in both parts of this work can be used to show that such a desired state can robustly emerge and persist against strategic deviations.


\appendix
\section{Proofs of Section~\ref{sec:local_ev_stability}}

\subsection{Proof of Theorem~\ref{th:imitative_rest_unstable}}\label{sec:proof_imitative_rest_unstable}

\begin{lemma}\label{lem:imitative_monotone_growth_rates}
	Consider an imitative revision protocol $\rho^c$. Then, the evolutionary dynamics \eqref{eq:ODE_mu_ev} follow $\dot{\mu}^c[\Scal^c,u] = \mu^c[\Scal^c,u]G^c_u(\mu)$, where
	\begin{equation}\label{eq:imitative_growth_rate}
		\begin{split}
		G^c_u(\mu) = \sum_{u^\prime \in \Ucal^c_D} \frac{\mu^c[\Scal^c,u^\prime]}{m^c}\Big(&r^c_{u^\prime u}(F^c(\mu),\mu^c[\Scal^c,\cdot])\\ &-r^c_{u u^\prime}(F^c(\mu),\mu^c[\Scal^c,\cdot])\Big)
	\end{split}
	\end{equation}
	is called a growth rate. Furthermore, the growth rates are monotonous, i.e., $F^c_u(\mu) \geq F^c_v(\mu) \iff G^c_u(\mu) \geq G^c_v(\mu) \;\forall c\in[C]\; \forall u,v\in \Ucal_D^c$, and $\sum_{u\in\Ucal_D^c}\mu^c[\Scal^c,u]G^c_u(\mu) = 0 \; \forall c\in[C]$. 
\end{lemma}
\begin{proof}
	One can write the evolutionary dynamics of the total mass on each policy by summing \eqref{eq:ODE_mu_ev} over $\Scal^c$ as
	\begin{equation}\label{eq:ODE_mu_ev_proof_sum_S}
		\begin{split}
		\dot{\mu}^c[\Scal^c,u] =  &\sum_{u^\prime \in \Ucal_D^c}  \mu^c[\Scal^c,u^\prime] \rho^c_{u^\prime u}(F^c(\mu),\mu^c[\Scal^c,\cdot])  \\&- \mu^c[\Scal^c,u] \sum_{u^\prime \in \Ucal_D^c}  \rho^c_{uu^\prime}(F^c(\mu),\mu^c[\Scal^c,\cdot])
		\end{split}
	\end{equation}
	for all $u\in \Ucal^c_D$ since by conservation of mass $\sum_{s\in \Scal^c} f^{c,d}_{s,u}(\mu) = 0$ for all $u\in \Ucal^c_D$. Using the definition of imitative revision protocol in \cite[Definition~7]{PedrosoAgazziEtAl2025MFGAvg} in \eqref{eq:ODE_mu_ev_proof_sum_S} yields 
	\begin{equation*}
			\begin{split}
			\dot{\mu}^c[\Scal^c,u] & = \!\! \sum_{u^\prime \in \Ucal^c_D} \!\! \mu^c[\Scal^c,u^\prime] r^c_{u^\prime u}(F^c(\mu),\mu^c[\Scal^c,\cdot])\mu^c[\Scal^c,u]/m^c \\
			& - \mu^c[\Scal^c,u]\!\! \sum_{u^\prime \in \Ucal^c_D} \!\! r^c_{uu^\prime}(F^c(\mu),\mu^c[\Scal^c,\cdot])\mu^c[\Scal^c,u^\prime]/m^c\\
			& = \mu^c[\Scal^c,u] \sum_{u^\prime \in \Ucal_D^c}  \frac{\mu^c[\Scal^c,u^\prime]}{m^c} \Big( r^c_{u^\prime u}(F^c(\mu),\mu[\Scal^c,\cdot]) \\
			& \quad \quad \quad \quad \quad \quad \quad \quad \quad \quad -  r^c_{uu^\prime}(F^c(\mu),\mu^c[\Scal^c,\cdot])\Big),\\
			& =  \mu^c[\Scal^c,u]G^c_u(\mu)
		\end{split}
	\end{equation*}
	for all $u\in \Ucal^c_D$ using the growth rates defined in \eqref{eq:imitative_growth_rate}. The monotonicity of the growth rates follows immediately from \cite[Definition~7]{PedrosoAgazziEtAl2025MFGAvg} and \eqref{eq:imitative_growth_rate}. The equality $\sum_{u\in\Ucal^c_D}\mu[\Scal^c,u]G_u^c(\mu) = 0$ follows from mass preservation, i.e., $ 0 = \sum_{u\in\Ucal^c_D}	\dot{\mu}^c[\Scal^c,u] = \sum_{u\in\Ucal^c_D}  \mu^c[\Scal,u]G^c_u(\mu)$.%
\end{proof}

\begin{lemma}\label{lem:imitative_nonMSNE_rest}
	Consider that at least one class $c\in [C]$ uses an imitative via comparison revision protocol, while the others use excess payoff or pairwise comparison revision protocols. Let $\mu^\star$ be a non-MSNE rest-point of \eqref{eq:ODE_mu_ev}. Then, there exists one class $d\in [C]$ that uses an imitative via comparison revision protocol for which optimal policies are not in the support of $\mu^{d\star}$ and their growth rates are positive, i.e., $v\in \argmax_{j \in \Ucal^d_D}  F^d_j(\mu^\star) \implies ( \mu^{d\star}[\Scal^d,v] = 0 \;\land\; G^d_v(\mu^\star) > 0)$.
\end{lemma}
\begin{proof}
	Since $\mu^\star$ is by hypothesis a rest point of \eqref{eq:ODE_mu_ev}, by \cite[Theorem~5]{PedrosoAgazziEtAl2025MFGAvg} it follows that classes that use use excess payoff or pairwise comparison revision protocols satisfy both conditions of the definition of a MSNE in \cite[Definition~2]{PedrosoAgazziEtAl2025MFGAvg}. Since, by hypothesis, $\mu^\star$ is not a MSNE, then by definition of MSNE, there is a class $d\in [C]$ that uses an imitative via comparison revision protocol for which at least one of the following statements is false
	\begin{enumerate}[(i)]
		\item For all $u \in \Ucal^d_D$, $\mu^{d\star}[\Scal^d,u] >0 \implies F^d_u(\mu^\star) \geq F^d_{v}(\mu^\star)$ for all $v\in\Ucal^d_D$;
		\item For all $s\in \Scal^d$ and $u\in \Ucal^c_D$, $f^{d,d}_{s,u}(\mu^\star) = 0$.
	\end{enumerate}
	First, we proceed to prove that statement~(i) is always false. Assume, by contradiction, that statement~(i) is true, i.e., all the mass is placed on optimal policies. As a result, since  $\mu^\star$ is not a MSNE, statement~(ii) must be false. Since statement~(i) is true we are in the conditions of \cite[Lemma~7]{PedrosoAgazziEtAl2025MFGAvg}, thus it follows from \cite[Lemma~7(ii)]{PedrosoAgazziEtAl2025MFGAvg} that $f^{d,r}_{s,u}(\mu^\star) = 0$ for all $s\in \Scal^d$ and all $u\in \Ucal_D^d$. Since $\mu^\star$ is a rest point of \eqref{eq:ODE_mu_ev}, then $f^{d,d}_{s,u}(\mu^\star)+ f^{d,r}_{s,u}(\mu^\star) = 0$ for all $s\in \Scal^d$ and $u\in \Ucal_D^d$. One then concludes that $f^{d,d}_{s,u}(\mu^\star) = 0$ for all $s\in \Scal^d$ and all $u\in \Ucal_D^d$, which is a contradiction with the fact that statement~(ii) must be false. Second, since we just proved that statement~(i) is false, it follows that the exist $u,v^\prime \in \Ucal^d_D$ such that $\mu^{d\star}[\Scal^d,u]>0$ and $F^d_u(\mu^\star) <F^d_{v^\prime}(\mu^\star)$. Since any $v\in \argmax_{j \in \Ucal^d_D}  F^d_j(\mu^\star)$ satisfies $F^d_{v^\prime}(\mu^\star) \leq  F^d_v(\mu^\star)$, then  $F^d_u(\mu^\star) <F^d_{v}(\mu^\star)$. Moreover, since $\mu^\star$ is a rest point, $0 = \sum_{s\in \Scal^d} f^{d,r}_{s,u}(\mu^\star) +f^{d,d}_{s,u}(\mu^\star) = \sum_{s\in \Scal^d} f^{d,r}_{s,u}(\mu^\star) = \dot{\mu}[\Scal^d,u]$. Therefore, since $\mu^{d\star}[\Scal^d,u]>0$, it follows from Lemma~\ref{lem:imitative_monotone_growth_rates} that $G^d_u(\mu^\star) = 0$. By the monotonicity of the growth rates established in Lemma~\ref{lem:imitative_monotone_growth_rates}, it follows that $G^d_v(\mu^\star) >0$ since $F^d_v(\mu^\star) > F^d_u(\mu^\star)$. Again, since $\sum_{s\in \Scal^c} f^{d,r}_{s,v}(\mu^\star) = 0$ because $\mu^\star$ is a rest point, one concludes from Lemma~\ref{lem:imitative_monotone_growth_rates} that $\mu^{d\star}[\Scal^d,v] = 0$.
\end{proof}

By Lemma~\ref{lem:imitative_nonMSNE_rest}, it follows that there is a class $d\in [C]$ that uses an imitative via comparison revision protocol for which there is a policy $v\in \argmax_{j \in \Ucal^d_D}  F^d_j(\mu^\star)$ such that $\mu^{d\star}[\Scal^d,v] = 0$ and $G^d_v(\mu^\star) >0$. By continuity of the growth rates defined in Lemma~\ref{lem:imitative_monotone_growth_rates}, in a sufficiently small neighborhood of $\mu^\star$, denoted by $\Ocal_{\mu^\star}$, $G^d_v(\mu) \geq  k > 0$ for all $\mu \in \Ocal_{\mu^\star}$. Consider a solution trajectory $\{\mu(t)\}_{t\geq 0}$ of \eqref{eq:ODE_mu_ev} with $\mu(0)\in \Ocal_{\mu^\star}$. Notice that,  from Lemma~\ref{lem:imitative_monotone_growth_rates}, for all $\mu\in \Ocal_{\mu^\star}$,  $\dot{\mu}^d[\Scal^d,v](t) = \mu^d[\Scal^d,v](t)G^d_v(\mu(t))  \geq k  \mu^d[\Scal^d,v](t)$.
Since $k>0$, no matter how small $\Ocal_{\mu^\star}$ is, there is always $\mu(0) \in \Ocal_{\mu^\star}$ such that $\mu^d[\Scal^d,v](0) >0$. Therefore, $\mu^\star$ is not a Lyapunov stable equilibrium of \eqref{eq:ODE_mu_ev}. Furthermore, by Lemma~\ref{lem:imitative_monotone_growth_rates} and Lipschitz continuity of the vector field of the ODE, it follows that for any initial condition in $X$, the signal of the components of $\mu^d[\Scal^d,\cdot](t)$ is preserved forward in time. As a result, no solution trajectory of \eqref{eq:ODE_mu_ev} with initial condition in $\mathop{\mathrm{int}}(X)$ converges to $\mu^\star$.

\subsection{Proof of Theorem~\ref{th:stability_strict}}\label{sec:proof_stability_strict}

The proof relies on LaSalle's invariance principle~\cite[Chap.~4.2]{Khalil2002}. For the sake of simplicity, by a slight abuse of notation, given a trajectory $\mu(t) \in X$ we denote $\mu^c[\cdot,u](t):= \col(\mu^c[s,u](t),s\in \Scal^c) \in X^c_{\Ucal_D}$. By abuse of notation, we also write $f^{c,r}_{\cdot,u}(\mu) = \col(f^{c,r}_{s,u}(\mu), s\in \Scal^c)$, whose concatenation order is consistent with the one of $\mu^c[\cdot,u](t)$. Since the candidate Lyapunov function is not continuously differentiable, we use a generalized derivative called the upper right Dini derivative, which we define in what follows alongside with useful properties.

\begin{definition}[$\!\!\!${\cite[Definition A.15]{Bullo2024}}]
	The upper right Dini derivative of a continuous function $f:] a, b[\rightarrow \mathbb{R}$ at a point $t \in] a, b[$ is defined as
	\begin{equation*}
		D^{+} f(t)=\limsup _{\Delta t>0, \Delta t \rightarrow 0} \frac{f(t+\Delta t)-f(t)}{\Delta t}. \tag*{$\triangle$}
	\end{equation*}
\end{definition}

\begin{lemma}[$\!\!${\cite[Lemma A.16]{Bullo2024}}]\label{lemma:prop_Dini}
	Given a continuous function $f:] a, b[\rightarrow \mathbb{R}$
	\begin{enumerate}[(i)]%
		\item  if $f$ is differentiable at $t \in ] a, b[$, then $D^{+} f(t)=\frac{d}{d t} f(t)$ is the usual derivative of $f$ at $t$;
		\item  if $D^{+} f(t) \leq 0$  for all $t \in ] a, b[$, then $f$ is non-increasing on $] a, b[$.
	\end{enumerate}
\end{lemma}

Since $\mu^\star$ is a strict MSNE, by Definition~\ref{def:strict_MSNE}, for each class $c\in[C]$ there is a policy $u^{c\star} \in \Ucal^c_D$ such that $\mu^\star[\Scal^c,u^{c\star}] = m^c$ and $F^c_{u^{c\star}}(\mu^\star) > F^c_{v}(\mu^\star)$ for all $v\in \Ucal^c_D \setminus \{u^{c\star}\}$. We define a neighborhood $D_\alpha(\mu^\star) \subset X$ of $\mu^\star$ for some $\alpha >0$ as 
\begin{equation*}
	D_\alpha(\mu^\star)  := \left\{\mu\in  X : ||\mu-\mu^\star||_1 \leq \alpha\right\}.
\end{equation*} 
Before defining a candidate Lyapunov function, two propositions that allow to check the conditions of LaSalle's Theorem \cite[Theorem~A.7]{Bullo2024} are established.


\begin{proposition}\label{prop:strict_propoerties_rev}
	Consider an imitative, separable excess payoff, or pairwise comparison revision protocol for each class $c\in [C]$. Then, for all $c\in [C]$, there exist $\alpha^c_1,k^c\geq 0$ such that for all $\mu\in D_{\alpha_1^c}(\mu^\star)$
	\begin{enumerate}[(i)]
		\item $k^c\dot{\mu}^c[\Scal^c,u^{c\star}] \geq  ||f^{c,r}_{\cdot,u^{c\star}}(\mu)||_1  \geq 0$;
		\item $\dot{\mu}^c[\Scal^c,u^{c\star}]= 0 \iff \sum_{u\in \Ucal^c_D\setminus \{u^{c\star}\}}\mu^c[\Scal^c,u] = 0$.
	\end{enumerate}
\end{proposition}
\begin{proof}
First, we consider any class $c$ that uses an imitative revision protocol. By Lemma~\ref{lem:imitative_monotone_growth_rates}, it follows that $G^c_{u^{c\star}}(\mu^\star)\mu^{c\star}[\Scal^c,u^{c\star}] = 0$. Therefore, since $\mu^c[\Scal^c,u^{c\star}] >0$, $G^c_{u^{c\star}}(\mu^\star) = 0$. By the monotonicity of the growth rates established in Lemma~\ref{lem:imitative_monotone_growth_rates}, it follows that $G^c_{u}(\mu^\star)<0$  for all $u \in \Ucal_D^c \setminus \{u^{c\star}\}$. Therefore, by continuity, there is $\alpha>0$ such that $ \mu \in D_{\alpha}(\mu^\star) \implies G^c_{u}(\mu)<0 \;\forall c\in [C]\; \forall u \in \Ucal_D^c \setminus \{u^{c\star}\}$. By Lemma~\ref{lem:imitative_monotone_growth_rates}, $\sum_{u\in \Ucal^c_D}G^c_{u}(\mu)\mu^c[\Scal^c,u] = 0$, therefore 
\begin{equation}\label{eq:aux_expand_sum_G}
	G^c_{u^{c\star}}(\mu)\mu^c[\Scal^c,u^{c\star}] = - \sum_{u\in \Ucal_D^c \setminus \{u^{c\star}\}} G^c_{u}(\mu)\mu^c[\Scal^c,u].
\end{equation}
Since $G^c_{u}(\mu^\star)<0$  for all $u \in \Ucal_D^c \setminus \{u^{c\star}\}$,  it follows from \eqref{eq:aux_expand_sum_G} that $G^c_{u^{c\star}}(\mu)\geq 0$ for all $\mu \in D_{\alpha}(\mu^\star)$. Also from \eqref{eq:aux_expand_sum_G} one concludes that
\begin{equation}\label{eq:aux_proof_strict_imitative_onec}
	\begin{split}
	\dot{\mu}^c[\Scal^c,u^{c\star}] & = G^c_{u^{c\star}}(\mu)\mu^c[\Scal^c,u^{c\star}] \\& =  -\sum_{u\in \Ucal_D^c \setminus \{u^{c\star}\}} G^c_u(\mu)\mu^c[\Scal^c,u].
	\end{split}
\end{equation}
Again, since $G^c_{u}(\mu^\star)<0$ for all $u \in \Ucal_D^c \setminus \{u^{c\star}\}$, from \eqref{eq:aux_proof_strict_imitative_onec} one notices that $\dot{\mu}^c[\Scal^c,u^{c\star}] = 0$ if and only if $\sum_{u\in \Ucal_D^c \setminus \{u^{c\star}\}}\mu^c[\Scal^c,u] = 0$, thereby showing statement~(ii) for class $c$. From \eqref{eq:aux_proof_strict_imitative_onec}, for all $\mu\in D_{\alpha}(\mu^\star)$ and all $c\in[C]$
\begin{equation}\label{eq:aux_bound_k1}
	\begin{split}
	&\;\sum_{u\in \Ucal_D^c \setminus \{u^{c\star}\}} \mu^c[\Scal^c,u] \\
	\leq &\;  \dot{\mu}^c[\Scal^c,u^{c\star}]/\min_{u\in \Ucal_D^c \setminus \{u^{c\star}\}}\{-G^c_u(\mu)\}\\ \leq &\;k^c_1\dot{\mu}^c[\Scal^c,u^{c\star}],
	\end{split}
\end{equation}
where $k^c_1 \!>\!0$ is the maximum of $1/\min_{u\in \Ucal_D^c \setminus \{u^{c\star}\}} \{-G^c_u(\mu)\}$ in $\mu\in D_{\alpha}(\mu^\star)$, which exists since it is the minimum of a continuous function in a compact set. From~\eqref{eq:fd_fr}, it follows that $f^{c,r}_{s,u^{c\star}}(\mu) \leq \sum_{u\in \Ucal_D^c \setminus \{u^{c\star}\}}\mu^c[s,u]\rho^c_{uu^\star}(F^c(\mu),\mu^c[\Scal^c,\cdot])$, which can be written as $f^{c,r}_{s,u^{c\star}}(\mu) \leq k_2^c\sum_{u\in \Ucal_D^c \setminus \{u^{c\star}\}}\mu^c[s,u] \leq k^c_2 \sum_{u\in \Ucal_D^c \setminus \{u^{c\star}\}}\mu^c[\Scal^c,u]  $, where $k_2^c \geq 0$ is the maximum of $\rho^c_{uu^\star}(F^c(\mu),\mu^c[\Scal^c,\cdot])$ over $D_{\alpha}(\mu^\star)$, which exists since it is the maximum of a continuous function in a compact set. From  \eqref{eq:aux_proof_strict_imitative_onec}, $\sum_{s\in \Scal^c}f^{c,r}_{s,^{c\star}}(\mu) = \sum_{s\in \Scal^c}f^{c,d}_{s,u^{c\star}}(\mu)+ f^{c,r}_{s,u^{c\star}}(\mu)  = 	\dot{\mu}^c[\Scal^c,u^{c\star}]  \geq 0$ for all $\mu\in D_{\alpha}(\mu^\star)$. Since $f^{c,r}_{s,u^{c\star}}(\mu) \leq k_2^c \sum_{u\in \Ucal_D^c \setminus \{u^{c\star}\}}\mu^c[\Scal^c,u]$, then for all $s\in \Scal^c$ and all $\mu\in D_{\alpha}(\mu^\star)$, $f^{c,r}_{s,u^\star}(\mu) \geq - k^c_2(p^c-1) \sum_{u\in \Ucal_D^c \setminus \{u^{c\star}\}}\mu^c[\Scal^c,u]$. One concludes that
\begin{equation*}
	||f^{c,r}_{\cdot,u^\star}(\mu)||_1  \leq k_2^cp^c(p^c-1) \!\!\!\!\!\!\!\sum_{u\in \Ucal_D^c \setminus \{u^{c\star}\}}\!\!\!\!\!\!\!\mu^c[\Scal^c,u] \leq k^c\dot{\mu}^c[\Scal^c,u^\star],
\end{equation*}
where the last inequality follows from \eqref{eq:aux_bound_k1} and $k^c = k^c_1 k^c_2p^c(p^c-1)$, which proves statement~(i) for class $c$.

Second, we consider any class $c$ that uses a pairwise comparison protocol. Since $\mu\in D_{\alpha}(\mu^\star)$ is a bounded neighborhood of $\mu^\star$ and  $F^c_{u^\star}(\mu^\star) > F^c_{u}(\mu^\star)$ for all $u\in \Ucal_D^c \setminus \{u^{c\star}\}$, by continuity, for a sufficiently small $\alpha$, $F^c_{u^{c\star}}(\mu) > F^c_{u}(\mu)$ for all $u\in \Ucal_D^c \setminus \{u^{c\star}\}$ and all $\mu\in D_{\alpha}(\mu^\star)$. Therefore, by the definition of a pairwise comparison revision protocol, $\rho^c_{u^{c\star} u}(F^c(\mu),\mu^c[\Scal^c,\cdot]) = 0$ and $\rho^c_{uu^{c\star} }(F^c(\mu),\mu^c[\Scal^c,\cdot]) > 0$  for all $u\in \Ucal_D^c \setminus \{u^{c\star}\}$ and all $\mu\in D_{\alpha}(\mu^\star)$. As a result, 
\begin{equation}\label{eq:pairwise_cmp_signa_fr_strict}
	f_{s,u^\star}^{c,r}(\mu) = \!\!\!\!\!\!\! \sum_{u\in \Ucal_D^c \setminus \{u^{c\star}\}}\!\!\!\!\!\!\! \mu^c[s,u]\rho_{uu^{c\star} }(F^c(\mu),\mu^c[\Scal^c,\cdot]) \geq 0.
\end{equation}
Since $f_{s,u^{c\star}}^{c,r}(\mu) \geq 0$, $\dot{\mu}^c[\Scal^c,u^\star] =  ||f^{c,r}_{\cdot,u^{c\star}}(\mu)||_1$, statement~(i) holds with $k^c = 1$ for class $c$. Furthermore, for any $s\in \Scal^c$, \eqref{eq:pairwise_cmp_signa_fr_strict} holds with equality if and only if $\sum_{u\in \Ucal_D^c \setminus \{u^{c\star}\}}\mu^c[s,u] = 0$, which establishes statement~(ii) for class $c$.

Third, we consider any class $c$ that uses a separable excess payoff revision protocol. From the expression for the average payoff, it follows that for $u\in \Ucal_D^c \setminus \{u^{c\star}\}$ and $\mu \in D_\alpha(\mu^\star)$
\begin{align}\label{eq:aux_avg_payoff}
		\hat{F}^c_u(\mu) &= F^c_u(\mu)-\sum_{v\in \Ucal^c_D}\mu^c[\Scal^c,v]F^c_v(\mu)/m^c\notag \\
		& = F^c_u(\mu) - \mu^c[\Scal^c,u^{c\star}]F^c_{u^{c\star}}(\mu)/m^c \notag\\
		& \quad \quad \quad   - \sum_{v\in \Ucal_D^c \setminus \{u^{c\star}\}}\mu^c[\Scal^c,v]F^c_v(\mu)/m^c\notag\\
		& =  -(F^c_{u^{c\star}}(\mu) -F^c_u(\mu))\notag \\
		&  \quad \quad \quad   +  \sum_{v\in \Ucal_D^c \setminus \{u^{c\star}\}}\mu^c[\Scal^c,v](F^c_{u^{c\star}}(\mu) -F^c_v(\mu))/m^c\notag\\
		& \leq - \min_{v\neq u^{c\star}}\{F^c_{u^{c\star}}(\mu) -F^c_v(\mu)\} \\
		&  \quad \quad \quad+ \alpha \max_{v\in \Ucal_D^c \setminus \{u^{c\star}\}}\{F^c_{u^{c\star}}(\mu) -F^c_v(\mu)\}/m^c.\notag
\end{align}
Notice that, similarly to the continuity argument put forward in the proof for a pairwise comparison  revision protocol, for sufficiently small $\alpha^\prime>0$, there exists $k^c_3>0$ such that $\min_{v\in \Ucal_D^c \setminus \{u^{c\star}\}}\{F^c_{u^{c\star}}(\mu) -F^c_v(\mu)\} > k_3^c$ for all $\mu \in D_{\alpha^\prime}(\mu^\star)$. Similarly, there is $k^c_4>0$ such that  $\max_{v\in \Ucal_D^c \setminus \{u^{c\star}\}}\{F^c_{u^{c\star}}(\mu) -F^c_v(\mu)\} < k^c_4$ for all $\mu \in D_{\alpha^\prime}(\mu^\star)$ because it is the maximum of a continuous function in a compact set. It follows from \eqref{eq:aux_avg_payoff} that $\hat{F}^c_u(\mu)< -k^c_3+\alpha k^c_4$ for all $u \in \Ucal_D^c \setminus \{u^{c\star}\}$ and all $\mu \in D_{\alpha^\prime}(\mu^\star)$. Therefore, choosing sufficiently small $\alpha$ such that $0< \alpha < \alpha^\prime$, it follows that $\hat{F}^c_u(\mu)< 0$ for all $u \in \Ucal_D^c \setminus \{u^{c\star}\}$ and all $\mu \in D_\alpha(\mu^\star)$. Since the excess payoff protocol is separable, it is also sign preserving  (see \cite[Exercise~5.5.6]{Sandholm2010}), i.e., $\sign(\tau^c_u(\hat{F})) = \sign(\max(0,\hat{F}^c_u(\mu)))$, therefore $\tau^c_u(\hat{F}^c(\mu))= 0$ for all $u \in \Ucal_D^c \setminus \{u^{c\star}\}$ and all $\mu \in D_\alpha(\mu^\star)$. As a result, $f_{s,u^{c\star}}^{c,r}(\mu) = \tau^c_{u^{c\star}}(\hat{F}^c(\mu))\sum_{u \in \Ucal_D^c \setminus \{u^{c\star}\}} \mu^c[s,u] \geq 0$.
Since $f_{s,u^{c\star}}^{c,r}(\mu) \geq 0$, $\dot{\mu}^c[\Scal^c,u^{c\star}] =  ||f^{c,r}_{\cdot,u^{c\star}}(\mu)||_1$, therefore statement~(i) holds with $k^c = 1$ for class $c$. Furthermore, $\hat{F}^c_{u^{c\star}}(\mu) \geq 0$ for all  $\mu \in D_\alpha(\mu^\star)$ with equality if and only if $\mu^c[\Scal^c,u^{c\star}] = m^c$, therefore $\tau^c_u(\hat{F}^c(\mu)) \geq 0$ with equality if and only if $\mu^c[\Scal^c,u^{c\star}] = m^c$.  As a result, $\dot{\mu}^c[\Scal^c,u^{c\star}]=  \tau_{u^{c\star}}(\hat{F}^c(\mu))\sum_{u \in \Ucal_D^c \setminus \{u^{c\star}\}} \mu^c[\Scal^c,u]$ is null if and only if  $\mu^c[\Scal^c,u^{c\star}] = m^c$, which establishes statement~(ii) for class $c$.
\end{proof}

\begin{proposition}\label{prop:dini_state_strict}
	If $\{\mu(t)\}_{t\geq0}$ is a solution trajectory to \eqref{eq:ODE_mu_ev}, then $D^+||\mu^c[\cdot,u^{c\star}](t)-\mu^{c\star}[\cdot,u^{c\star}]||_1 \leq ||f^{c,r}_{\cdot,u^{c\star}}(\mu(t))||_1\;$ for all $t\geq 0$ and all $c\in [C]$. 
\end{proposition}
\begin{proof}
	For any $c\in [C]$, from \eqref{eq:ODE_mu_ev} one can write
	\begin{equation}\label{eq:strict_linear_sys_with_rev_input}
		\dot{\mu}^c[\cdot,u^{c\star}](t) = Q^{c,u^{c\star}} \mu^c[\cdot,u^{c\star}](t) + f^{c,r}_{\cdot,u^{c\star}}(\mu(t))
	\end{equation}
	where $Q^{c,u^{c\star}}$ is the generator of the Markov chain whose transition kernel is $\phi^{c,u^{c\star}} : \Scal \to \Pcal(\Scal)$ defined by $\phi^{c,u^{c\star}}(s|s^\prime) =  \sum_{a^\prime \in \Acal^c(s^\prime)}\phi^c(s|s^\prime,a^\prime)u^{c\star}(a^\prime|s^\prime)$. Analyzing \eqref{eq:strict_linear_sys_with_rev_input} as a linear system with an exogenous input $f^{c,r}_{\cdot, u^{c\star}}(\mu(t))$, one may conclude that it is weakly infinitesimally contracting~\cite[Definition~4.2]{Bullo2024}. Therefore, comparing the solution trajectory $\{\mu(t)\}_{t\geq0}$ with the degenerate solution trajectory that remains at $\mu^{c\star}$ and using \cite[Theorem~3.16]{Bullo2024}\footnote{Although \cite[Theorem~3.16]{Bullo2024} is stated for strongly contracting systems the proof can be reused with minimal changes to establish an analogous result for weakly contracting systems.} one obtains $D^+||\mu^c[\cdot,u^{c\star}](t)-\mu^{c\star}[\cdot,u^{c\star}]||_1 \leq ||f^{c,r}_{\cdot,u^{c\star}}(\mu(t))||_1$ for all $t\geq 0$.
\end{proof}

Let $\{\mu(t)\}_{t\geq0}$ be a solution trajectory to \eqref{eq:ODE_mu_ev}. The candidate Lyapunov function is 
\begin{equation}\label{eq:lyapunov_strict}
	\begin{split}
	V(\mu(t)) &=  \sum_{c\in [C]} ||\mu^c[\cdot,u^{c\star}](t)-\mu^{c\star}[\cdot,u^{c\star}]||_1\\ &+ \sum_{c\in [C]} C^c\!\!\!\!\! \sum_{v \in \Ucal_D^c \setminus \{u^{c\star}\}}||\mu^c[\cdot,v](t) - \mu^{c\star}[\cdot,v]||_1,
	\end{split}
\end{equation}
where $C^c>0$ with $c\in[C]$ is a constant to be chosen.
By the subadditivity of the $\limsup$ operator, it follows that $D^+V(\mu(t)) \leq \sum_{c\in [C]}D^+ ||\mu^c[\cdot,u^{c\star}](t)-\mu^{c\star}[\cdot,u^{c\star}]||_1 + \sum_{c\in[C]}C^cD^+\sum_{v \in \Ucal_D^c \setminus \{u^{c\star}\}}||\mu^c[\cdot,v](t)||_1$. Since the second term in \eqref{eq:lyapunov_strict} is differentiable, it follows from statement~(i) of Lemma~\ref{lemma:prop_Dini} that its upper right Dini derivative is the usual derivative, so \begin{equation*}
	\begin{split}
	&\sum_{c\in[C]}C^cD^+ \sum_{v \in \Ucal_D^c \setminus \{u^{c\star}\}}||\mu^c[\cdot,v](t)||_1 \\
	= &\sum_{c\in [C]} C^c \sum_{v \in \Ucal_D^c \setminus \{u^{c\star}\}} \dot{\mu}^c[\Scal^c,v](t) = -\sum_{c\in [C]}C^c \dot{\mu}[\Scal^c,u^{c\star}](t) 
	\end{split}
\end{equation*}
where the last equality is due to $\sum_{v\in\Ucal_D^c} \dot{\mu}[\Scal^c,v](t)  = 0$. As a result, from Proposition~\ref{prop:strict_propoerties_rev}(i) and Proposition~\ref{prop:dini_state_strict}, it follows that choosing $C^c>k^c$ for all $c\in [C]$
\begin{equation}\label{eq:Dplus_ineq}
	\begin{split}
	\!\!D^+V(\mu(t)) &\leq  \!\!\!\sum_{c\in [C]}\!||f^{r,c}_{\cdot,u^{c\star}}(\mu(t))||_1 \! -\!\!\!\sum_{c\in[C]}\!\!C^c\dot{\mu}^c[\Scal^c,u^{c\star}](t)\! \!\\
	 &\leq   -\sum_{c\in[C]}(C^c-k^c)\dot{\mu}^c[\Scal^c,u^{c\star}](t)  \leq 0,
	\end{split}
\end{equation}
for all $\mu(t) \in D_{\alpha_1}(\mu^\star)$, where $\alpha_1 = \min_{c\in [C]}\alpha^c_1$.
Notice that the set $\Omega := \left\{\mu \in X : V(\mu) \leq \alpha_2 \right\}$ is compact for any $\alpha_2$ and, for sufficiently small $\alpha_2>0$,  $\Omega \subset D_{\alpha_1}(\mu^\star)$. As a result, $D^+V(\mu(t)) \leq 0$ if $\mu(t) \in \Omega$ and, by statement~(ii) of Lemma~\ref{lemma:prop_Dini}, $\Omega$ is a positively invariant set. By a simple generalization of LaSalle's Invariance Principle \cite[Theorem~3.16]{Bullo2024} every trajectory approaches the largest invariant set in $E = \{\mu \in \Omega : D^+V(\mu) = 0\}$. For details on the generalization, one can refer to the discussion on \cite[Chap.~A.7]{Bullo2024} or check that the proof of LaSalle's Invariance Principle in \cite[Theorem~4.4]{Khalil2002} holds in this setting as well. One can conclude that
\begin{equation*}
	\begin{split}
	D^+V(\mu) = 0 &\implies \sum_{c\in[C]}(C^c-k^c)\dot{\mu}^c[\Scal^c,u^{c\star}] = 0 \\
	&\implies  \dot{\mu}^c[\Scal^c,u^{c\star}] = 0 \; \forall c\in [C] \\
	& \implies \mu^c[\Scal^c,u^{c\star}] = m^c\; \forall c\in [C],
	\end{split}
\end{equation*}
where the first implication follows from \eqref{eq:Dplus_ineq}, the second from  Proposition~\ref{prop:strict_propoerties_rev}(i), and the third from Proposition~\ref{prop:strict_propoerties_rev}(ii). Therefore, by \cite[Assumption~2]{PedrosoAgazziEtAl2025MFGAvg}, the largest invariant set contained in $E$ is  $\{\mu^\star\}$, which allows to conclude that $\mu^\star$ is locally asymptotically stable.

\subsection{Proof of Lemma~\ref{lem:ODE_x_z}}\label{sec:proof_ODE_x_z}

The following properties of $S^c$ and $\bar{S}^c$ are instrumental for this proof, as well as the proof of Theorem~\ref{th:tts_stability_generic}.

\begin{proposition}\label{prop:prop_S}
	The linear transformations $\BSK^c, \BKS^c, \BSbarK^c$, and $\BKSbar^c$ follow: (i)~$\ones^\top \BSK^c = \ones^\top$; (ii)~$\ones^\top\BSbarK^c = \zeros$;  (iii)~$\BSK^c \BKS^c + \BSbarK^c \BKSbar^c= I$; (iv)~$Q^c \BSK^c = \zeros$; (v)~$\BKSbar^c \BSK^c = \zeros$; (vi)~$\BKSbar^c \BSbarK^c  = I$; (vii)~$\BKSbar^c Q^c \BSbarK^c$ is Hurwitz; (viii)~$\BKS^c (I-Q^c\BSbarK^c (\BKSbar^c Q^c \BSbarK^c)^{-1}  \BKSbar^c) = \BKS^c = I_{n^c} \otimes \ones_{p^c}^\top$; (ix) $\BKS^c Q^c = \zeros$.
\end{proposition}
\begin{proof}
	Statements (i)-(vii) follow immediately from how these transformations are defined. To show statement~(viii), by abuse of notation, we say, for $x^c\in \C^{n^c}$, that $x^c\in S^c$ if $\Re(x^c)\in S^c$ and $\Im(x^c)\in S^c$; and for $z^c\in \C^{(p^c-1)n^c}$, that $z^c\in \bar{S}^c$ if $\Re(z)\in \bar{S}^c$ and $\Im(z)\in \bar{S}^c$.  Let $\sigma \in \C^{p^cn^c}$ be an eigenvector of $Q^c$ associated with eigenvalue $\lambda$. Notice that exactly $n^c$ eigenvalues are null, by \cite[Assumption~2]{PedrosoAgazziEtAl2025MFGAvg}, which are associated with an eigenvector $\sigma \in S^c$. Henceforth, let $\lambda$ be nonnull. Since $Q^c$ generates a Markov chain, then $\lambda < 0$, and $\BKSbar^c \sigma \neq \zeros$. By the definition of an eigenvalue eigenvector pair, it follows that $Q^c\sigma = \lambda \sigma$. By statement~(iii), it follows that $x^c = \BKS^c \sigma \in S^c$ and $z^c = \BKSbar^c \sigma \in \bar{S}^c$ satisfy $\sigma = \BSK^c x^c + \BSbarK^c z^c$. Moreover, $z^c\neq \zeros$. Therefore, by statements~(iv)-(vi), it follows that $\BKSbar^c Q^c \BSbarK^c z^c = \lambda z^c$ and $z^c\neq \zeros$. One may then conclude that the $(p^c-1)n^c$ negative eigenvalues of $Q^c$ are all the eigenvalues of $\BKSbar^c Q^c \BSbarK^c$ and are associated with eigenvectors  $\BKSbar^c \sigma$. Therefore, $\BKSbar^c Q^c \BSbarK^c$ is Hurwitz. To prove statement~(viii), we first notice that, by definition, $\BKS^c$ is characterized by $\BKS^c \BSK^c = I$ and $\BKS^c \BSbarK^c = \zeros$. Since $\BKS^c =   I_{n^c} \otimes \ones_{p^c}^\top$ abides by both, then it is the closed-form expression for $\BKS^c$. Since $(I_{n^c} \otimes \ones_{p^c}^\top )Q^c = \zeros$, it follows immediately that $\BKS^c (I-Q^c\BSbarK^c (\BKSbar^c Q^c \BSbarK^c)^{-1}  \BKSbar^c) = \BKS^c = ( I_{n^c} \otimes \ones_{p^c}^\top)$. Statement~(ix) follows from statement~(viii), since $\BKS^c Q^c = (I_{n^c} \otimes \ones_{p^c}^\top)Q^c = \zeros$.%
\end{proof}

Making use of \eqref{eq:ODE_mu_ev} and Propositions~\ref{prop:prop_S}(iv) and~\ref{prop:prop_S}(ix), the time-evolution of the coordinates $x(t) = \BKS \mu(t)$ and $z(t) = \BKSbar \mu(t)$ is described by
	\begin{equation}\label{eq:aux_proof_x_z_dyn}
	\begin{split}
		\dot{x}^c(t) &=  \BKS^c f^{c,r}(\BSK x + \BSbarK z) \\
		\epsilon \dot{z}^c(t) &= \frac{\Rdc}{\Rdbar}\BKSbar^c Q^c \BSbarK^c z^c \\
		& \quad \quad \quad \quad  + \epsilon \BKSbar f^{c,r}(\BSK x + \BSbarK z),
	\end{split}
\end{equation}
for all $c\in[C]$. From \eqref{eq:aux_proof_x_z_dyn} and Proposition~\ref{prop:prop_S}(viii), it follows that $\dot{x}^c_u(t) = \sum_{s\in \Scal^c}f^{c,r}_{s,u}(\BSK x + \BSbarK z)$ for all $c\in [C]$ and all $u\in \Ucal_D^c$. Expanding this equation using \eqref{eq:fd_fr} yields the $x$ component of \eqref{eq:standard_singular_perturbation}. The $z$ component of \eqref{eq:standard_singular_perturbation} follows immediately from \eqref{eq:aux_proof_x_z_dyn}. Furthermore, \eqref{eq:aux_proof_x_z_dyn} enjoys the same regularity properties as \eqref{eq:ODE_mu_ev}, as described in the proof of \cite[Lemma~5]{PedrosoAgazziEtAl2025MFGAvg}, therefore its solution exists in $t\in [0,\infty)$ and is unique. Therefore, by the definition of the change of basis matrices, $\mu(t) = \BSK x(t) + \BSbarK z(t)$ is a solution to \eqref{eq:ODE_mu_ev}. Furthermore, since $\ones^\top \dot{x}^c(t) = \sum_{s\in \Scal^c}\sum_{u\in \Ucal_D^c}f^{c,r}_{s,u}(\BSK x + \BSbarK z) = 0$ and $\ones^\top x^c(0) = \ones^\top \BKS^c \mu^c(0) = m^c$ by Proposition~\ref{prop:prop_S}(viii), then $\ones^\top x(t) =m^c$ for all $t\geq 0$, which concludes the proof of the lemma.

\subsection{Proof of Theorem~\ref{th:tts_stability_generic}}\label{sec:proof_tts_stability_generic}

We start by defining, in the following proposition, a Lyapunov function for the \emph{boundary-layer} system
\begin{equation}\label{eq:tts_bd_layer_sys}
	\epsilon \dot{z}(t) = \BKSbar \bar{Q} \BSbarK z,
\end{equation}
where $\bar{Q}:= \diag(\Rdc Q^c, c\in [C])/\Rdbar$.

\begin{proposition}\label{prop:lyapunov_bd_layer}
	There exists a Lyapunov function $W:D_z \to \Rnn$ of the boundary-layer system \eqref{eq:tts_bd_layer_sys} which satisfies: (i)~$W$ is continuously differentiable; (ii)~$W$ is positive definite, i.e., $W(z) > 0$ for all $z\in D_z\setminus \{\zeros\}$ and  $W(\zeros) = 0$; (iii)~$\dot{W}(z) \leq \gamma_3 ||z||^2$ for all $z\in D_z$ for some $\gamma_3>0$; and (iv)~$||\partial W /\partial z|| \leq \gamma_4 ||z||$ for some $\gamma_4>0$.
\end{proposition}
\begin{proof}
	Notice that $\bar{Q}$ is a block diagonal matrix with blocks $\BKSbar^c Q^c \BSbarK^c\Rdc/\Rdbar$ for $c\in[C]$. Since, by Proposition~\ref{prop:prop_S}(vii), $\BKSbar^c Q^c \BSbarK^c$ is Hurwitz for all $c\in [C]$, then $\bar{Q}$ is Hurwitz. Hence, the origin of the boundary-layer system \eqref{eq:tts_bd_layer_sys} is globally exponentially stable. The result follows immediately from Lyapunov's converse theorem \cite[Theorem~4.14]{Khalil2002}.
\end{proof}

To establish asymptotic stability of $\{(x,z): x\in \Mcal_{\Fcal},\; z = \zeros\}$ under \eqref{eq:standard_singular_perturbation} in the two time-scale regime, we study a candidate Lyapunov function $\nu : \bar{D}_x\times D_z \to \Rnn$ which is the weighted sum of the Lyapunov functions for the reduced system \eqref{eq:tts_red_sys} in the conditions of the theorem and for the boundary-layer system in Proposition~\ref{prop:lyapunov_bd_layer}, i.e., 
\begin{equation*}
	\nu(x,z) = (1-\theta)V(x) + \theta W(z),
\end{equation*}
where $\theta\in (0,1)$ is a parameter to be chosen. In what follows, we establish properties of $\nu$ that allow to apply Lyapunov's stability theorems.

First, since  both $V$ and $W$ are continuously differentiable, then $\nu$ is continuously differentiable. Second, since both $V$ and $W$ are positive definite, then $\nu(x,z) = 0$ if and only if $x\in \Mcal_\Fcal$ and $z =  \zeros$, therefore $\nu$ is positive definite. Third, the time derivative of $\nu$ along trajectories of \eqref{eq:standard_singular_perturbation} is given by
\begin{equation}\label{eq:nu_dot_big}
	\begin{split}
		\!\!\!\!&\;\dot{\nu}(x,z)  =  (1-\theta)\frac{\partial V}{\partial x}\BKS f^r(\BSK x) \\
		\!\!\!\!&+ \! (1\!-\!\theta)\frac{\partial V}{\partial x}\BKS(f^r\!(\BSK x \!+\! \BSbarK z) \!-\! f^r\!(\BSK x)\!)\!\!\!\!\!\! \\
		\!\!\!\!& + \frac{\theta}{\epsilon} \frac{\partial W}{\partial z}\BKSbar \bar{Q} \BSbarK z \\
		\!\!\!\!& +  \theta\frac{\partial W}{\partial z} \BKSbar f^r(\BSK x + \BSbarK z).
	\end{split}
\end{equation}
We proceed to upper bound each of the terms in \eqref{eq:nu_dot_big}. By the upper bound on $\dot{V}(x)$ in the conditions of the theorem
\begin{equation}\label{eq:nu_dot_big_1}
	(1-\theta)\frac{\partial V}{\partial x}\BKS f^r(\BSK x) \leq -(1-\theta) \gamma_1 d_{\Mcal_\Fcal}^2(x).
\end{equation}
By \cite[Assumptions~1 and~3]{PedrosoAgazziEtAl2025MFGAvg}, $\BKS f^r(\BSK x + \BSbarK z)$ is Lipschitz continuous w.r.t.\ $z$ uniformly in $x$, therefore there is a uniform Lipschitz constant $L_z^S>0$ such that $||\BSK f^r(\BSK x + \BSbarK z) - \BSK f^r(\BSK x)|| \leq  L_z^S||z||$. Additionally, by the bound on $||\partial V /\partial x||$ in the conditions of the theorem, it follows that
\begin{align}\label{eq:nu_dot_big_2}
		&(1-\theta)\frac{\partial V}{\partial x}\BKS(f^r(\BSK x + \BSbarK z) \!-\! f^r(\BSK x)) \notag \\
	\leq & (1-\theta)\gamma_2d_{\Mcal_\Fcal}(x) L_z^S ||z||.
\end{align}
From Proposition~\ref{prop:lyapunov_bd_layer}(iii) it follows that
\begin{equation}\label{eq:nu_dot_big_3}
	 \frac{\theta}{\epsilon} \frac{\partial W}{\partial z}\BKSbar \bar{Q} \BSbarK z  \leq - \frac{\theta}{\epsilon}\gamma_3||z||^2.
\end{equation}
Finally, since from condition~(i) of the theorem $f^r(\BSK x^\star) = \zeros$ for all $x^\star \in \Mcal_\Fcal$, then $ f^r(\BSK x + \BSbarK z) = f^r(\BSK x + \BSbarK z) - f^r(\BSK x) +  f^r(\BSK x)- f^r(\BSK x^\star)$. By \cite[Assumptions~1 and~3]{PedrosoAgazziEtAl2025MFGAvg}, $f^r(\BSK x + \BSbarK z)$ is Lipschitz continuous w.r.t.\ $z$ uniformly in $x$ and w.r.t.\ $x$ uniformly in $z$, therefore there exist Lipschitz constants $L_x,L_z^{\bar{S}}>0$ such that $||\BKSbar(f^r(\BSK x + \BSbarK z) - f^r(\BSK x))|| \leq L_s^{\bar{S}}||z||$ and $||\BKSbar (f^r(\BSK x)- f^r(\BSK x^\star))|| \leq L_x ||x-x^\star|| = L_x d_{\Mcal_\Fcal}(x)$ for an appropriate choice of $x^\star \in \Mcal_\Fcal$. As a result, also from Proposition~\ref{prop:lyapunov_bd_layer}(iv),
\begin{equation}\label{eq:nu_dot_big_4}
	\begin{split}
	&\theta\frac{\partial W}{\partial z} \BKSbar f^r(\BSK x + \BSbarK z) \\
	\leq& \theta\gamma_4||z||(L_xd_{\Mcal_\Fcal}(x) +L_z^{\bar{S}}||z||).
	\end{split}
\end{equation}
\begin{figure*}[ht]
	\begin{equation}\label{eq:nu_dot_too_long}
		\begin{split}
			\dot{\nu}(x,z) & \leq -(1-\theta) \gamma_1 d_{\Mcal_\Fcal}^2(x) - \theta(\gamma_3/\epsilon - \gamma_4L_z^{\bar{S}})||z||^2 +((1-\theta)\gamma_2 L_z^S +\theta\gamma_4L_x)||z||d_{\Mcal_\Fcal}(x)\\
			& \leq -\begin{bmatrix}
				d_{\Mcal_\Fcal}(x)\\ ||z||
			\end{bmatrix}^\top
			\begin{bmatrix}
				(1-\theta) \gamma_1 & -\frac{1}{2}(1-\theta)\gamma_2 L_z^S -\frac{1}{2}\theta\gamma_4L_x \\ -\frac{1}{2}(1-\theta)\gamma_2 L_z^S -\frac{1}{2}\theta\gamma_4L_x & \theta(\gamma_3/\epsilon - \gamma_4L_z^{\bar{S}})
			\end{bmatrix} \begin{bmatrix}
				d_{\Mcal_\Fcal}(x)\\ ||z||
			\end{bmatrix}.
		\end{split}
	\end{equation}
\end{figure*}
From, \eqref{eq:nu_dot_big_1}, \eqref{eq:nu_dot_big_2}, \eqref{eq:nu_dot_big_3}, and \eqref{eq:nu_dot_big_4}, one can upper bound $\dot{\nu}(x,z)$ by \eqref{eq:nu_dot_too_long} below.
Let
\begin{equation*}
	\epsilon^\star := \frac{\theta(1-\theta)\gamma_1\gamma_3}{\frac{1}{4}((1-\theta)\gamma_2 L_z^S -\frac{1}{2}\theta\gamma_4L_x)^2 + \theta(1-\theta)\gamma_1\gamma_4L_z^{\bar{S}}},
\end{equation*}
where $\theta\in (0,1)$ can be chosen such that $\epsilon^\star$ is as large as possible. Since $\Mcal_\Fcal$ is closed by hypothesis, it follows that, for all $\epsilon < \epsilon^\star$, $\dot{\nu}(x,z) < 0$ for all $(x,z) \in \bar{D}_x \times D_z \setminus \{(x,z): x\in \Mcal_{\Fcal},\; z = \zeros\}$ and  $\dot{\nu}(x,z) = 0$ for all $(x,z) \in \{(x,z): x\in \Mcal_{\Fcal},\; z = \zeros\}$. One can conclude that, for all $\epsilon < \epsilon^\star$, $\nu$ is a Lyapunov function for \eqref{eq:standard_singular_perturbation} and by \cite[Theorem~4.1]{Khalil2002} $\{(x,z): x\in \Mcal_{\Fcal},\; z = \zeros\}$ is locally asymptotically stable under  \eqref{eq:standard_singular_perturbation}.

\subsection{Proof of Theorem~\ref{th:tts_ultimate_bd}}\label{sec:proof_tts_ultimate_bd}

In the following proposition, we start by establishing an ultimate boundedness result on the $z$ component of the evolutionary dynamics of the dynamic game in \eqref{eq:standard_singular_perturbation}. Intuitively, given $B_z >0$, there is a sufficiently small $\epsilon$ such that $z(t)$ reaches $||z(t)||\leq B_z$ in finite time and satisfies it for all future time. 

\begin{proposition}\label{prop:ultimate_bd_layer}
	There exist $\gamma_5>0$ and $\gamma_6 \in (0,1]$ such that for all $B_z >0$ and for all $\epsilon <  \epsilon^\star_z:= \gamma_5 B_z$: (i)~there is a finite $T_{B_z} >0$ such that 
	\begin{equation*}
		 ||z(t)|| \leq B_z, \quad \forall t\geq T_{B_z} \quad \forall (x(0),z(0)) \in D_x\times D_z;
	\end{equation*}
	and (ii)~for all $x(0) \in D_x$ 
	\begin{equation*}
	 ||z(0)||\leq \gamma_6 B_z \implies ||z(t)||\leq \gamma_6 B_z \;\; \forall t\geq 0.
	\end{equation*}
	\begin{proof}
		By \cite[Assumptions~1 and~3]{PedrosoAgazziEtAl2025MFGAvg}, $f^r(\BSK x + \BSbarK z)$ is continuous w.r.t.\ $x$ and $z$ and since it is defined on a compact domain $D_x\times D_z$, then there is $B_f^r >0$ such that $||\BKSbar f^r(\BSK x + \BSbarK z)||\leq B_f^r$ for all $(x,z) \in D_x\times D_z$. Therefore, one can write the dynamics of the $z$ component of \eqref{eq:standard_singular_perturbation} as 
		\begin{equation}\label{eq:standard_singular_perturbation_z_dist}
			\dot{z}(t) = \frac{1}{\epsilon} \BKSbar \bar{Q} \BSbarK z(t) + d(t),
		\end{equation}
		where $\bar{Q}:= \diag(\Rdc Q^c, c\in [C])/\Rdbar$ and $d(t)$ is regarded as a time-varying disturbance that satisfies $||d(t)||\leq B_{f^r}$ for all $t\geq 0$. Since, by Proposition~\ref{prop:prop_S}(vii), $\BKSbar Q^c \BSbarK$ is Hurwitz, then $\BKSbar \bar{Q} \BSbarK$ is Hurwitz and, from \cite[Theorem~4.6]{Khalil2002}, given any real positive definite matrix $R$, there is a unique real symmetric positive definite solution $P$ to the Lyapunov equation $P  \BKSbar \bar{Q} \BSbarK + ( \BKSbar \bar{Q} \BSbarK)^\top = -R$. Define $W:D_z \to \Rnn$ as $W(z) = z^\top Pz$. Notice that $W$ satisfies the following properties: (i)~$W$ is continuously differentiable; (ii)~lower and upper bounded by
		\begin{equation}\label{eq:W_bounds}
			||z||^2\lambdamin(P) \leq W(z) \leq \lambdamax(P) ||z||^2,
		\end{equation}
		where $\lambdamin(P)>0$ and $\lambdamax(P)>0$ denote the smallest and largest eigenvalue of $P$, respectively; and (iii)~the time derivative of $W$ along trajectories of \eqref{eq:standard_singular_perturbation_z_dist} satisfies 
		\begin{equation}\label{eq:W_dot_init}
			\begin{split}
				\dot{W}(z) &= \frac{\partial W}{\partial z}\dot{z}(t) = \frac{1}{\epsilon}z^\top (PA+A^\top P)z + 2z^\top P d(t)\\
				& \leq -\frac{1}{\epsilon}\lambdamin(R)||z||^2 + 2||z||\lambdamax(P)B_{f^r}.
			\end{split}
		\end{equation}
		Given any $B_z>0$, notice that defining $\epsilon^\star_z := \lambdamin(R)\sqrt{\lambdamin(P)/\lambdamax(P)}B_z/(2 \lambdamax(P)B_{f^r})$, one can rewrite the bound on $\dot{W}(z)$ in \eqref{eq:W_dot_init} as
		\begin{equation*}
			\begin{split}
				\dot{W}(z) & \leq -\left(\frac{1}{\epsilon}-\frac{1}{\epsilon_z^\star}\right)\lambdamin(R)||z||^2 \\
				&+ 2\lambdamax(P)B_{f^r}||z||\left(1-\frac{||z||}{\sqrt{\lambdamin(P)/\lambdamax(P)}B_z}\right)\!.
			\end{split}
		\end{equation*}
		One may then conclude that
		\begin{equation*}
			\dot{W}(z) \leq -\left(\frac{1}{\epsilon}-\frac{1}{\epsilon^\star}\right)\lambdamin(R)||z||^2
		\end{equation*}
		$\forall ||z||\geq\sqrt{\lambdamin(P)/\lambdamax(P)}B_z$. From \eqref{eq:W_bounds}, it follows that $||z||\leq \sqrt{\lambdamin(P)/\lambdamax(P)}B_z \implies z\in \Omega := \{ z\in D_z: W(z)\leq \lambdamin(P)B_z^2\}$. Therefore, since $\dot{W}(z)$ is negative  for all $\epsilon > \epsilon^\star_z$ and for all $z\in D_z \setminus \mathrm{int}\,\Omega$, it follows that $\Omega$ is positively invariant and any trajectory starting in $D_z$ enters $\Omega$ in finite time $T_{B_z} > 0$ and stays therein for all $t\geq T_{B_z}$ (see \cite[Theorem~4.18]{Khalil2002} for a generic treatment of such Lyapunov arguments). Since, by \eqref{eq:W_bounds}, $z\in \Omega \implies ||z|| \leq B_z$, one can conclude that any trajectory starting in $D_z$ satisfies $||z(t)||\leq B_z$ for all $t\geq T_{B_z}$. Furthermore, if $||z(0)|| \leq \sqrt{\lambdamin(P)/\lambdamax(P)}B_z$, i.e., $||z(0)||\in \Omega$, then $z(t)\in \Omega$ for all $t\geq 0$ because $\Omega$ is positively invariant.
	\end{proof}
\end{proposition}

The concept of a class $\Kcal$ is instrumental for the remainder of the proof and is defined in what follows.

\begin{definition}[{\cite[Definition~4.2]{Khalil2002}}]
	A continuous function $\alpha: [0, a) \to [0,\infty)$ is said to belong to class $\Kcal$ if it is strictly increasing and $\alpha(0) = 0$. \hfill $\triangle$
\end{definition}

By \cite[Assumptions~1 and~3]{PedrosoAgazziEtAl2025MFGAvg}, $f^r(\BSK x + \BSbarK z)$ is Lipschitz continuous w.r.t.\ $z$  uniformly in $x$, therefore there is a Lipschitz constant $L$ such that $||\BKS( f^r(\BSK x + \BSbarK z) - f^r(\BSK x))||\leq L||z||$ for all $(x,z) \in D_x\times D_z$. Therefore, one can write the dynamics of the $x$ component of \eqref{eq:standard_singular_perturbation} (which are written in a simpler form in \eqref{eq:aux_proof_x_z_dyn}) as
\begin{equation}\label{eq:standard_singular_perturbation_x_dist}
	\begin{split}
			\dot{x}(t) &= \BKS f^r(\BSK x + \BSK z)\\
			& =  \BKS f^r(\BSK x) + d(t),
	\end{split}
\end{equation}
where $d(t)$ is regarded as time-varying disturbance that satisfies $||d(t)||\leq L||z(t)||$ for all $t\geq 0$. By hypothesis, the Lyapunov function $V$ is defined on a neighborhood of $\Mcal_\Fcal$ denoted by $\bar{D}_x$, therefore there is $B_x>0$ such that $d_{\Mcal_\Fcal}(x)\leq B_x \implies x\in \bar{D}_x$. By hypothesis, there exists $B_V >0$ such that $||\partial V/\partial x|| \leq B_V$ for all $x\in \bar{D}_x$. Since $V$ is positive definite and $\dot{V}$ is negative definite, by \cite[Lemma~4.3]{Khalil2002}, there exist class $\Kcal$ functions $\alpha_1,\alpha_2 $, and $\alpha_3$ defined on $[0,B_x]$ such that 
\begin{equation*}
	\alpha_1(d_{\Mcal_\Fcal}(x)) \leq V(x) \leq \alpha_2(d_{\Mcal_\Fcal}(x))
\end{equation*}
and 
\begin{equation*}
	\frac{\partial V}{\partial x}\BKS f^r(\BSK x) \leq -\alpha_3(d_{\Mcal_\Fcal}(x)).
\end{equation*}
When $d_{\Mcal_\Fcal}(x)\leq B_x$, the evolution of $V$ along trajectories of \eqref{eq:standard_singular_perturbation_x_dist} is given by 
\begin{equation}\label{eq:Vdot_x}
	\begin{split}
		\dot{V}(x) &= \frac{\partial V}{\partial x}\BKS f^r(\BSK x) + \frac{\partial V}{\partial x}d(t)\\
		& \leq -\alpha_3(d_{\Mcal_\Fcal}(x)) + B_VL||z||.
	\end{split}
\end{equation}

In what follows, we prove the local result of the theorem resorting to Proposition~\ref{prop:ultimate_bd_layer}(ii). Given any $B>0$, one may choose $\theta \in (0,1)$ and
$B_z>0$ sufficiently small such that $d_{\Mcal_\Fcal}(x) \leq \alpha_1^{-1}(\alpha_2(\alpha_3^{-1}(B_VL\gamma_6B_z/\theta))) < B_x$ and $||z||\leq \gamma_6B_z$ imply $d_\Mcal(\mu) \leq B$. When $d_{\Mcal_\Fcal}(x)\leq B_x$ and if $||z(0)|| \leq \gamma_6B_z$, one can rewrite \eqref{eq:Vdot_x} for all $\epsilon < \gamma_5B_z$ by Proposition~\ref{prop:ultimate_bd_layer}(ii) as
\begin{equation*}
		\begin{split}
		\dot{V}(x) &\!\leq \!-\alpha_3(d_{\Mcal_\Fcal}(x)) + B_VL\gamma_6B_z\\
		& \!\leq \!-(1\!-\!\theta) \alpha_3(d_{\Mcal_\Fcal}(x)) \!-\! \big(\theta  \alpha_3(d_{\Mcal_\Fcal}(x))  \!-\!  B_VL\gamma_6B_z\big).
	\end{split}
\end{equation*}
Since $\theta$ and $B_z$ were chosen such such that $\alpha_3^{-1}(B_VL\gamma_6B_z/\theta) < \alpha_2^{-1}(\alpha_1(B_x))$ one can resort to an analysis similar to the proof of Proposition~\ref{prop:lyapunov_bd_layer}. Indeed, if $||z(0)|| \leq \gamma_6$, for all $\epsilon <  \gamma_5B_z$ and for all $d_{\Mcal_\Fcal}(x) \geq \alpha_3^{-1}(B_VL\gamma_6B_z/\theta)$ it follows that 	$\dot{V}(x) \leq -(1-\theta) \alpha_3(d_{\Mcal_\Fcal}(x))$, therefore if $d_{\Mcal_\Fcal}(x(0)) \leq \alpha_2^{-1}(\alpha_1(B_x)) $, then there exists finite $T\geq 0$ such that $d_{\Mcal_\Fcal}(x(t)) \leq \alpha_1^{-1}(\alpha_2(\alpha_3^{-1}(B_VL\gamma_6B_z/\theta)))$ for all $t\geq T$. Notice that there is $B_0$ such that $d_\Mcal(\mu) \leq B_0$ implies $d_{\Mcal_\Fcal}(x) \leq \alpha_2^{-1}(\alpha_1(B_x)) $ and $||z||\leq \gamma_6B_z$. One may conclude that for all $B>0$ for all $\epsilon > \epsilon^\star = \gamma_5B_z$ if $d_\Mcal(\mu) \leq B_0$ then $d_\Mcal(\mu) \leq B$ for all $t\geq T$.

The global result of the theorem, i.e., when $\bar{D}_x = D_x$, follows from a similar analysis of \eqref{eq:Vdot_x} resorting to Proposition~\ref{prop:ultimate_bd_layer}(i). For all $(x(0),z(0))\in D_x \times D_z$, it follows from Proposition~\ref{prop:ultimate_bd_layer}(i) that $||z(t)||\leq B_z$ for all $t\geq T_{B_z}$. Therefore, from  \eqref{eq:Vdot_x}, for $t\geq T_{B_z}$
\begin{equation}
	\begin{split}
		\dot{V}(x)  \leq -\alpha_3(d_{\Mcal_\Fcal}(x)) + B_VLB_z.
	\end{split}
\end{equation}
Making the same arguments as for the local result, for all $B>0$ one can choose $\theta \in (0,1)$ and $B_z$ such that for all $\epsilon > \epsilon^\star = \gamma_5B_z$ there is $T_x>0$ such that $d_\Mcal(\mu) \leq  B $ for all $t\geq T := T_{B_z}+T_x$ and for all $(x(0),z(0))\in D_x \times D_z$.


\subsection{Proof of Corollary~\ref{th:ub_potential_global}}\label{sec:proof_ub_potential_global}

Since a potential function $U$ for the static game $\Fcal$ exists, then $\NE(\Fcal)$ is the set of maximizers of $U$ in $D_x$ \cite[Theorem~3.1.3]{Sandholm2010}. Since $U$ is concave and continuous (because it is differentiable), then its set of maximizers, i.e., $\NE(\Fcal)$, is nonempty, compact, and convex. One can define the candidate Lyapunov function $V(x) = U(x^\star)-U(x)$, where $x^\star \in \NE(\Fcal)$ is chosen arbitrarily. Notice that $V$ is continuously differentiable and $||\partial V(x)/\partial x|| = ||\Fcal(x)||$ is bounded in $D_x$ since $\Fcal(x)$ is continuous by \cite[Assumption~1]{PedrosoAgazziEtAl2025MFGAvg} and $D_x$ is compact. Moreover, $V(x) \geq 0$ for all $x\in D_x$ and $V(x) = 0$ if and only if $x\in \NE(\Fcal)$. The time derivative of $V$ along trajectories of \eqref{eq:tts_red_sys} is given by $\dot{V}(x) = -\nabla U^\top \dot{x} = - \Fcal^\top(x)\dot{x} = -\sum_{c\in[C]} \Fcal^{c\top}(x)\dot{x}^c$.
By \cite[Theorem~5.5.2]{Sandholm2010} and \cite[Theorem~5.6.2]{Sandholm2010}, for any excess payoff or pairwise comparison revision protocol $\rho^c$, it follows that $\Fcal^{c\top}(x)\dot{x}^c \geq 0$ for all $x\in D_x$ with equality if and only if $\dot{x}^c = \zeros$. Therefore, it follows that if each class uses excess payoff or pairwise comparison revision protocols, then $\dot{V}(x) \leq 0$ with equality if and only if $\dot{x} = \zeros$, or equivalently by \cite[Theorem~5]{PedrosoAgazziEtAl2025MFGAvg}, $x \in \NE(\Fcal)$. Therefore, for excess payoff and pairwise comparison revision protocols we are in the conditions of the global statement of Theorem~\ref{th:tts_ultimate_bd}, which proves the result immediately. Now, we turn to the case where some classes may use imitative revision protocols with the initial condition $\mu(0) \in X^\star$. By \cite[Thereom~5.4.7]{Sandholm2010}, evolutionary imitative dynamics for $x^c(t)$ of a class $c\in \Ccal_\mathrm{I}$ are positively invariant in $X^{c\star} = \{x^c_0\in D^c_x: x^{c\star}(u) > 0  \implies x^c_0(u) > 0, \forall u\in \Ucal^c_D\; \forall x^\star \in \NE(\Fcal)\}$. For any imitative revision protocol, by \cite[Theorem~5.4.9]{Sandholm2010}, $\Fcal^{c\top}(x)\dot{x}^c \geq 0$ for all $x\in D_x$ and, by \cite[Theorem~5.4.13]{Sandholm2010}, when $x^c\in X^{c\star}$, $\Fcal^{c\top}(x)\dot{x}^c = 0$ if and only if $\dot{x}^c = \zeros$. Therefore, it follows that if each class uses imitative, excess payoff, or pairwise comparison revision protocols and $\mu(0) \in X^\star$, then $\sum_{c\in[C]} \Fcal^{c\top}(x)\dot{x}^c \geq 0$ with equality if and only if $\dot{x} = \zeros$. By the properties of excess payoff \cite[Theorem~5.5.2]{Sandholm2010}, pairwise comparison \cite[Theorem~5.6.2]{Sandholm2010}, and imitative revision protocols over $X^\star$ \cite[Theorem~5.4.13]{Sandholm2010},  $\dot{x} = \zeros$ if and only if $x\in \NE(\Fcal) \subseteq X^\star$. Therefore, defining the same Lyapunov function over $X^\star$ allows to apply the global statement of Theorem~\ref{th:tts_ultimate_bd} for imitative revision protocols.

\subsection{Proof of Lemma~\ref{lem:MF}}\label{sec:proof_lem_MF}

Statement~(i) follows immediately from the fact that $ \ker \Pi \cap \R_{\Ucal_D^\star}$ is a linear space and $D_x$ is compact. Statement~(ii) follows from the fact that $\Fcal(x)$ and $\Fcal(y)$ are a function  of $\Pi x$ and  $\Pi y$. Since $\Pi x = \Pi (x^\star + w_x) = \Pi x^\star =  \Pi (x^\star + w_y) =  \Pi y$, where $w_y,w_x \in  \ker \Pi \cap \R_{\Ucal_D^\star}$, then $\Fcal(x) = \Fcal(y)$. Statement~(iii) follows from the fact that $x^\star \in \NE(\Fcal)$, payoffs are constant in $\Mcal_\Fcal$ by statement~(ii), and mass is only placed in payoff maximizing policies since $x-x^\star \in \R_{\Ucal_D^\star}$ for all $x\in \Mcal_\Fcal$. Statement~(iv) follows from writing $\Fcal(x)$ as a function of $\Pi x$ as $\Fcal(\Pi x)$ and noticing that $D\Fcal(\Pi x) = D_{\Pi x}\Fcal(\Pi x)\Pi = D_{\Pi x}\Fcal(\Pi x^\star)\Pi = D_{\Pi x}\Fcal(\Pi y)\Pi = D\Fcal(\Pi y)$. 
 
\subsection{Proof of Proposition~\ref{prop:check_ESS}}\label{sec:proof_prop_check_ESS}
For all $x\in \Mcal_\Fcal$ and for all $w \in TX \cap \R_{\Ucal_D^\star}\setminus  \ker \Pi$, notice that $w = \Phi v$ for $v = \Phi^\top w$. Therefore, 
\begin{equation}\label{eq:aux1_ESS_diff_cond}
	\begin{split}
	w^\top D\Fcal(x) w &= v^\top \Phi^\top D\Fcal(x) \Phi v \\
	&= \frac{1}{2}v^\top \Phi^\top ( D\Fcal(x)+  D\Fcal(x)^\top ) \Phi v.
	\end{split}
\end{equation}
By the spectral theorem for symmetric matrices, there is an orthonormal basis of $\R^{n^\star-C}$ consisting of the eigenvectors of $\Phi^\top ( D\Fcal(x)+  D\Fcal(x)^\top ) \Phi$. Therefore, $w$ can be written as $w = \Phi v = \Phi (v_0 +v_<)$ where $v_0$ is an eigenvector associated with the null eigenvalue and $v_<$ is a linear combination of eigenvectors associated with nonnull eigenvalues such that $v = v_0 +v_<$. One can rewrite \eqref{eq:aux1_ESS_diff_cond} as
\begin{equation}\label{eq:aux2_ESS_diff_cond}
	\begin{split}
		&2w^\top D\Fcal(x) w \\
		&=  (v_0 +v_<)^\top\Phi^\top ( D\Fcal(x)+  D\Fcal(x)^\top ) \Phi   (v_0 +v_<)\\
		& = (v_0 +v_<)^\top\Phi^\top ( D\Fcal(x)+  D\Fcal(x)^\top ) \Phi   v_<\\
		& = v_<^\top \Phi^\top ( D\Fcal(x)+  D\Fcal(x)^\top ) \Phi   v_0 \\
		& \quad \quad \quad \quad \quad \quad  \quad \quad+ v_<^\top\Phi^\top ( D\Fcal(x)+  D\Fcal(x)^\top ) \Phi   v_< \\
		& =  v_<^\top\Phi^\top ( D\Fcal(x)+  D\Fcal(x)^\top ) \Phi   v_<.
	\end{split}
\end{equation}
Then \eqref{eq:ESS_diff_cond} holds, i.e., \eqref{eq:aux2_ESS_diff_cond} is negative for all $w \in TX \cap \R_{\Ucal_D^\star}\setminus  \ker \Pi$ if and only if all eigenvalues of $\Phi^\top ( D\Fcal(x)+  D\Fcal(x)^\top ) \Phi$ are nonpositive and every real eigenvector $v\in \R^{n^\star-C}$ in the eigenspace associated with the null eigenvalue satisfies $w = \Phi v  \in \ker \Pi$. Finally, notice that $G +G^\top = \Phi^\top ( D\Fcal(x)+  D\Fcal(x)^\top ) \Phi$.

\subsection{Proof of Lemma~\ref{lem:ESS}}\label{sec:proof_lem_ESS}

Let $\Mcal_\Fcal$ be a regular ESS and $\Ocal$ an arbitrarily small neighborhood of $\Mcal_\Fcal$. First, we prove that $\Mcal_\Fcal$ is isolated by contradiction.  Consider, by contradiction, that there is a NE $y\in \Ocal \setminus \Mcal_\Fcal$. Notice that since  $\Mcal_\Fcal$ is a regular ESS and the payoff is continuous and invariant in $\Mcal_\Fcal$ by Lemma~\ref{lem:MF}(ii), then $\Fcal^c_u(w) > \Fcal^c_v(w)$ for all $c\in[C]$, all $u\in \Ucal_D^{c\star}$, all $v \notin \Ucal_D^{c\star}$, and all $w\in \Ocal$. Therefore, if $y\in \Ocal$ is a NE, $y$ only places mass in policies in $\Ucal_D^{c\star}$ for each class $c\in [C]$. Since $\Ocal$ is an arbitrarily small neighborhood, there is $x\in \Mcal_\Fcal$ arbitrarily close to $y$. One concludes that $y-x \in \R_{\Ucal_D^\star}$, $y-x \in TX$, and $y-x \notin \ker \Pi$ (otherwise $y\in \Mcal_\Fcal$, which is not true by hypothesis). Therefore, by \eqref{eq:ESS_diff_cond} and Lemma~\ref{lem:MF}(iv), $(y-x)^\top D\Fcal(x) (y-x)< 0$. Performing a first order expansion of $\Fcal$ about $x$ yields $\Fcal(y) = \Fcal(x) + D\Fcal(x)(y-x) + \delta(||y-x||^2)$, where  $\delta(||y-x||^2)$ denotes higher order terms. Therefore, 
\begin{equation}\label{eq:isolated_1}
	\begin{split}
	&(y-x)^\top(\Fcal(y) -\Fcal(x)) \\
	= &(y-x)^\top D\Fcal(x)(y-x) + (y-x)^\top \delta(||y\!-\!x||^2) < 0 
	\end{split}
\end{equation}
since $y$ and $x$ can be made arbitrarily close. Moreover, since by hypothesis $y$ is a NE, for any $w\in \Ocal$ and any $c\in[C]$, $\sum_{u\in \Ucal^c_D} y^c_u\Fcal^c_u(y) \geq \sum_{u\in \Ucal^c_D} w^c_u\Fcal^c_u(y)$. Hence, $\sum_{c\in[C]}\sum_{u\in \Ucal^c_D} y^c_u\Fcal^c_u(y) \geq \sum_{c\in[C]}\sum_{u\in \Ucal^c_D} w^c_u\Fcal^c_u(y)$
thus $(y-w)^\top\Fcal(y)\geq 0$ for all  $w\in \Ocal$. Particularizing, $w = x$ yields $(y-x)^\top\Fcal(y)\geq 0$. Since $x\in  \Mcal_\Fcal$ is also a NE by Lemma~\ref{lem:MF}(iii), one can perform the same analysis to conclude that $(x-w)^\top\Fcal(x)\geq 0$ for all  $w\in \Ocal$. Particularizing for $w = y$ yields $(x-y)^\top\Fcal(x)\geq 0$. One concludes that 
\begin{equation}\label{eq:isolated_2}
	\begin{split}
	&(y-x)^\top(\Fcal(y)-\Fcal(x)) \\
	 = & (y-x)^\top\Fcal(y) +(x-y)^\top\Fcal(x)  \geq 0.
	\end{split}
\end{equation}
However, \eqref{eq:isolated_1} and \eqref{eq:isolated_2} lead to a contradiction, thereby proving the first part of the lemma. To prove the second statement of the lemma, we consider two cases: (i)~$w\in \ker \Pi$; and (ii)~$w\notin \ker \Pi$. Writing $\Fcal(x)$ as a function of $\Pi x$ as  $\Fcal(\Pi x)$ yields $D\Fcal(\Pi x)w = D_{\Pi x}\Fcal(\Pi x)\Pi w  =\zeros$ for all $x\in \Ocal$ and all $w\in \ker \Pi$. Thus, in case~(i), $w^\top D\Fcal(x) w = 0$ for all $x\in \Ocal$ and for all $w\in  TX \cap  \R_{\Ucal_D^\star} \cap \ker \Pi$. For case (ii), notice by the definition of ESS in \eqref{eq:ESS_diff_cond} that $w^\top D\Fcal(x^\star) w < 0$ for all $x^\star\in \Mcal_\Fcal$ and for all $w\in  TX \cap  \R_{\Ucal_D^\star}  \setminus \ker  \Pi$. For any $x \in \Ocal$ choosing $x^\star \in \argmin_{w\in \Mcal_\Fcal} ||x^\star-x||$ one can make $||x-x^\star||$ be arbitrarily small since the neighborhood $\Ocal$ can be made arbitrarily small. By the continuity of $D\Fcal$, which follows from \cite[Assumption~1]{PedrosoAgazziEtAl2025MFGAvg}, $w^\top D\Fcal(x) w < 0$ for all $x \in \Ocal$ and for all $w\in  TX \cap  \R_{\Ucal_D^\star}  \setminus \ker  \Pi$.



\section*{References}
\bibliographystyle{IEEEtran}
\bibliography{../../../../../Publications/bibliography/parsed-minimal/bibliography.bib,../../../../../Papers/_bib/references-gt.bib,../../../../../Papers/_bib/references-c.bib}

%

\end{document}